\documentclass[10pt,aps,pra,twocolumn,showpacs,superscriptaddress,nobalancelastpage,longbibliography]{revtex4-2}
\pdfoutput=1
\usepackage[utf8]{inputenc}
\usepackage[english]{babel}
\usepackage[T1]{fontenc}
\usepackage{amsmath}
\usepackage{dsfont}
\usepackage{enumitem}
\usepackage{hyperref}
\usepackage{graphicx}
\usepackage{dcolumn}
\usepackage{bm}
\usepackage{hyperref}
\usepackage[mathlines]{lineno}
\usepackage{amssymb}
\usepackage[f]{esvect}
\usepackage[braket, qm]{qcircuit}  
\usepackage{listings}

\usepackage{xcolor}
\usepackage{tabulary}
\usepackage{booktabs,tabularx}
\usepackage{amssymb,amsthm}
\usepackage{nicematrix}
\usepackage{relsize}
\usepackage{tikz}
\usepackage{lipsum}
\usepackage{import}
\usepackage{ulem}
\usepackage{algorithm,algpseudocode,algorithmicx}
\usepackage{braket}
\usepackage{cleveref}

\algnewcommand\algorithmicinput{\textbf{Input:}}
\algnewcommand\Input{\item[\algorithmicinput]}
\algnewcommand\algorithmicoutput{\textbf{Output:}}
\algnewcommand\Output{\item[\algorithmicoutput]}
\algnewcommand\algorithmiccomplexity{\textbf{Cost:}}
\algnewcommand\Complexity{\item[\algorithmiccomplexity]}
\algnewcommand\algorithmicparamter{\textbf{Parameter:}}
\algnewcommand\Parameter{\item[\algorithmicparamter]}

\newcommand{\red}[1]{\textcolor{red}{#1}}

\newcommand{\commentout}[1]{}

\newcommand{\delete}[1]{}

\DeclareMathOperator*{\argmin}{arg\,min}

\newcommand{\brm}[1]{\bm{\mathrm{#1}}}

\newcommand{\hl}{\brm{L}_{k}}
\newcommand{\hll}{\brm{L}_{k}^{\ell}}
\newcommand{\hlu}{\brm{L}_{k}^{u}}

\newcommand{\bk}{\brm{B}_{k}}
\newcommand{\bkdag}{\brm{B}_{k}^{\dagger}}
\newcommand{\bkk}{\brm{B}_{k+1}}
\newcommand{\bkkdag}{\brm{B}_{k+1}^{\dagger}}

\def\prob{\mathbb{P}}
\newcommand{\Prob}[1]{\mathbb{P}\left(#1\right)}
\newcommand{\CProb}[2]{\prob\left(\left.#1\right|#2\right)}
\newcommand{\Exp}[1]{\mathbb{E}\left[#1\right]}

\makeatletter
\newsavebox{\@brx}
\newcommand{\llangle}[1][]{\savebox{\@brx}{\(\m@th{#1\langle}\)}%
  \mathopen{\copy\@brx\kern-0.5\wd\@brx\usebox{\@brx}}}
\newcommand{\rrangle}[1][]{\savebox{\@brx}{\(\m@th{#1\rangle}\)}%
  \mathclose{\copy\@brx\kern-0.5\wd\@brx\usebox{\@brx}}}
\makeatother

\newtheorem{theorem}{Theorem}
\newtheorem{lemma}{Lemma}

\newtheorem{problem}{Problem}

\theoremstyle{remark}

\begin{document}

\title{Quantum HodgeRank: Topology-Based Rank Aggregation on Quantum Computers}

\author{Caesnan M. G. Leditto}
\email{caesnan.leditto@monash.edu}
\affiliation{School of Physics and Astronomy, Monash University, Clayton, VIC 3168, Australia}
\affiliation{Quantum Systems, Data61, CSIRO, Clayton, VIC 3168, Australia}

\author{Angus Southwell}
\affiliation{Quantum for New South Wales, Haymarket, NSW 2000, Australia}
\affiliation{School of Physics and Astronomy, Monash University, Clayton, VIC 3168, Australia}

\author{Behnam Tonekaboni}
\affiliation{Quantum Systems, Data61, CSIRO, Clayton, VIC 3168, Australia}

\author{Muhammad Usman}
\affiliation{Quantum Systems, Data61, CSIRO, Clayton, VIC 3168, Australia}
\affiliation{School of Physics, The University of Melbourne, Parkville, VIC 3052, Australia}
\affiliation{School of Physics and Astronomy, Monash University, Clayton, VIC 3168, Australia}

\author{Kavan Modi}
\email{kavan.modi@monash.edu}
\affiliation{Quantum for New South Wales, Haymarket, NSW 2000, Australia}
\affiliation{School of Physics and Astronomy, Monash University, Clayton, VIC 3168, Australia}

\begin{abstract}
HodgeRank generalizes ranking algorithms, e.g. Google’s PageRank, to rank alternatives based on real-world (often incomplete) data using graphs and discrete exterior calculus. It analyzes multipartite interactions on high-dimensional networks with a complexity that scales exponentially with dimension. We develop a quantum algorithm that approximates the HodgeRank solution with complexity independent of dimension. Our algorithm extracts relevant information from the state such as the ranking consistency, which achieves a superpolynomial speedup over similar classical methods.
\end{abstract}

\maketitle

\textit{Introduction.}---Ranking is crucial for informed decision-making and comparing alternatives. Google's PageRank~\cite{Brin1998PageRank,LangvilleMeyer+2006Google} and Netflix's recommendation system~\cite{Gomez2016Netflix} are both, at their core, ranking algorithms. Ranking systems typically operate on noisy or incomplete data. For instance, consider ranking football teams based on match results: the data compares the teams pairwise, but some pairs may be missing, and thus the data is incomplete. Such ranking data appears commonly in systems with many alternatives, as many pairs might not have been explicitly compared. 

HodgeRank~\cite{Jiang2011StatisticalTheory} overcomes such limitations by utilizing discrete exterior calculus~\cite{hiranithesis2003discrete,desbrun2005discrete} to aggregate incomplete data -- e.g., the pairwise match results -- to derive a global ranking. It also provides a consistency measure for its ranking, indicating whether the data implies a strong consensus ranking of alternatives or whether the data contains \textit{ranking inconsistencies}, such as Condorcet paradoxes (rock-paper-scissors scenarios). HodgeRank generalizes PageRank and has applications in recommendation systems, social choice theory, arbitrage-free market detection, sports tournament ranking, online video quality assessments, market analysis, and protein folding analysis~\cite{Jiang2011StatisticalTheory, xu2012hodgerank, Sizemore2014HodgeRank, Do2019Blockchain, Yang2022SimplicialFilters, Wei2022HodgeAnalysis}. 

The above examples involve ranking based on pairwise comparisons, but the HodgeRank tools of discrete exterior calculus on simplicial complexes generalize to higher-order networks~\cite{Battiston2020NetworksDynamics,Battiston2021TheSystems,Bianconi_2021,Majhi2022DynamicsReview,Bick2023WhatNetworks}. These are used to model complex phenomena such as biological networks~\cite{Haruna2016HodgeNetworks,Anand2023brain} and chaotic Hamiltonian dynamics~\cite{Millan2020ExplosiveComplexes,Carletti2023GlobalTopologicalSync}, with further applications to machine learning~\cite{Barbarossa2020TopologicalComplexes,Ebli2020SimplicialNetworks,Schaub2021SignalBeyond}. However, higher-dimensional HodgeRank ($k$-HodgeRank, for network dimension $k$) computations are expensive: the complexity scales exponentially in $k$. This hinders practical applications and exploration of higher-order network models despite strong evidence for their importance in real-world systems~\cite{Austin2016HigherOrderOrganization,Grilli2017HigherOrderCompetitive,Tekin2018HigherOrderDrug}. 

In this Letter, we propose a quantum algorithm that prepares a quantum state proportional to the output of $k$-HodgeRank that is exponentially faster in $k$ than generalizations of classical methods. This advantage is maintained when extracting answers to specific questions, modulo some caveats discussed later. We give algorithms to (i) measure rankability metrics of the data, (ii) relatively rank a subset of alternatives, and (iii) return top-ranking alternatives. In particular, the first applicationis superpolynomially faster for large $k$, and polynomially faster for $k=1$ (corresponding to pairwise preference data), than comparable classical methods. This is similar to other quantum algorithms that estimate the fit quality of data~\cite{Wiebe2012QDataFitting,Liu2017FastQLS}. The latter applications are more heuristic and data-dependent. While we focus on ranking algorithms here, the tools developed in the Letter will have broader applications to higher-dimensional discrete exterior calculus, such as solving general least squares~\cite{Ding2022HardnessGadgets} and boundary value problems~\cite{Schwarz_1995,Arnold_Falk_Winther_2006FEEC}. 


\textit{HodgeRank.}--- We briefly describe HodgeRank and the associated mathematical problems (see~\cite{Jiang2011StatisticalTheory} for details). HodgeRank outputs a global ranking of alternatives $\mathcal{V}:=\{v_{1}, \dots, v_{n}\}$.  The input to HodgeRank is aggregated pairwise preferences of voters, modelled by an edge flow $\brm{s}^{1}:\mathcal{E} \to \mathcal{R}$ on a comparison graph $\mathcal{G} = (\mathcal{V}, \mathcal{E})$. This corresponds to a vector $\brm{s}^1 \in \mathbb{R^{|\mathcal{E}|}}$ by (arbitrarily) orienting each edge in $\mathcal{E}$. The graph $\mathcal{G}$ need not be {complete: not all pairs of alternatives must be compared directly. The value $\brm{s}^{1}([v_i, v_j])$ represents the preference for alternative $i$ over $j$, where $[\cdot, \cdot]$ denotes edge orientation.

Given a comparison graph $\mathcal{G}$ and preference data $\brm{s}^{1}$, HodgeRank solves a weighted $\ell_{2}$ minimization problem with solution
\begin{eqnarray}\label{statistical ranking problem}
\brm{s}^1_{\mathrm{G}}=\argmin_{\brm{x} \in \mathrm{im}(\brm{B}_1^{\dagger})} \left[\sum_{1 \leq i < j \leq n} w_{ij}\left(x_{ij}-\brm{s}^{1}([v_{i},v_{j}])\right)^2\right],
\end{eqnarray}
where $w_{ij}$ represents the weight associated to the data on (oriented) edge $[v_i, v_j]$ and $\bk$ is a matrix representation of the $k$-th boundary operator of a clique complex $\mathcal{K}_{n}(\mathcal{G})$ (see Supplemental Material~\cite{supp}). The space $\mathrm{im}(\brm{B}_1^{\dagger})$ contains so-called \textit{conservative} edge flows, which means there exists $\brm{s}^0_* \in \mathbb{R}^{|\mathcal{V}|}$ such that 
$\brm{s}^1_\mathrm{G}([v_i, v_j]) = \brm{s}^0_*(v_i) - \brm{s}^0_*(v_j)$. In a discrete exterior calculus sense, $\brm{s}^1_\mathrm{G}$ is an analogue of the gradient of the \textit{potential function} $\brm{s}^0_*$. 

HodgeRank then outputs the potential (or score) function $\brm{s}^0_*$ from Eq.~\eqref{statistical ranking problem} by $\brm{B}_{1}^{\dagger}\brm{s}^{0}_{\ast} = \brm{s}^1_\mathrm{G}$. This defines a global ranking of the alternatives: the score of alternative $v_i$ is $\brm{s}^0_*(v_i)$, and $v_i$ outranks $v_j$ whenever $\brm{s}^0_*(v_i) \geq \brm{s}^0_*(v_j)$. In this Letter, we set $w_{ij} = 0$ if $v_i, v_j$ are not compared, and 1 otherwise; we discuss this assumption in Supplemental Material~\cite{supp}. Under this assumption, $\brm{s}^{0}_{\ast} = \big(\brm{B}_{1}\brm{B}_{1}^{\dagger}\big)^{+}\brm{B}_{1}\brm{s}^{1}$, where $+$ denotes the Moore-Penrose pseudoinverse. The relative length of $\brm{s}^1_\mathrm{G}$ compared to $\brm{s}^1$ measure the rankability of the alternatives according to $\brm{s}^1$.

In higher dimensions, $k$-HodgeRank generalizes the comparison graph to a clique complex $\mathcal{K}_{n}(\mathcal{G})$ of dimension $k$ defined on $\mathcal{V}$. 
Instead of $\brm{s}^1 \in \mathbb{R}^{|\mathcal{E}|}$ assigning a real number to each edge ($1$-simplex), the data $\brm{s}^k \in \mathbb{R}^{n_k}$ assigns a real number $s^{k}_{i}$ to each $k$-simplex $\sigma_k^{i}$ (a $(k+1)$-element subset of $\mathcal{V}$), where $i \in [n_k]:=\{0,\cdots,n_k-1\}$ and $n_k$ is the number of $k$-simplices in $\mathcal{K}_n(\mathcal{G})$. 
The solution to $k$-HodgeRank, the higher-dimensional analogue of Eq.~\eqref{statistical ranking problem}, is 
\begin{eqnarray}\label{higher HodgeRank solution}
\brm{s}^{k-1}_{\ast}=\left(\bk\bk^{\dagger}\right)^{+}\brm{B}_{k}\brm{s}^{k}.
\end{eqnarray}
This vector assigns a score to each $(k-1)$-simplex in $\mathcal{K}_{n}(\mathcal{G})$. Similarly we define $\brm{s}^k_\mathrm{G} := \brm{B}_k^\dagger\brm{s}^{k-1}_*$, the projection of $\brm{s}^k$ onto the space of conservative flows on the $k$-simplices of $\mathcal{K}_{n}(\mathcal{G})$. We define $k$-HodgeRank as finding $\brm{s}^{k-1}_*$ given the following (classically stored) information: (i) $\brm{s}^k$, (ii) a list of simplices in the clique complex $\mathcal{K}_n(\mathcal{G})$ (which includes the edges of $\mathcal{G}$).

\textit{Quantum least squares algorithms.}---The HodgeRank problem in Eq.~\eqref{statistical ranking problem} is a type of least squares problem. For $k=1$, $k$-HodgeRank can be solved exactly with complexity $O(n^3)$~\cite{Jiang2011StatisticalTheory}. 
Approximate algorithms, using graph Laplacian solvers~\cite{Spielman2004NearlyLinearTime,Spielmanm2014NearlySystems,Cohen2014Solving} or topological signal processing (TSP)~\cite{Barbarossa2020TopologicalComplexes,Yang2022SimplicialFilters}, have complexities approximately quadratic in $n$, with varying dependence on the approximation parameters.
For $k>1$, computing $\brm{s}^{k}_{\mathrm{G}}$ and $\brm{s}^{k-1}_*$ via least squares using existing classical sparse linear equation solvers has worst-case scaling $\Omega(n_k)$~\cite{Black2022HodgeComplexes,Ding2022HardnessGadgets}, since $\brm{B}_k$ is an $n_k \times n_{k-1}$ matrix with $(k+1)n_k$ non-zero entries. Generalizing the TSP approach similarly leads to algorithms with complexity $\Omega(n_k)$~\cite{Yang2021FiniteComplexes}. Since $n_k$ can grow like $\binom{n}{k+1}$, this can lead to exponential scaling in $k$.

There are several quantum algorithms for least square problems~\cite{Wiebe2012QDataFitting,Liu2017FastQLS,chakraborty_et_al2018PowerBlockEncoding,Kerenidis2017QuantumSystem,Chakraborty2023quantumregularized}. The best known algorithm requires $\tilde{O}(\alpha_{\brm{A}}\kappa_{\brm{A}}\log(1/\varepsilon))$ calls to a block encoding of $\brm{A}$, where $\alpha_{\brm{A}}$ is the scaling factor coming from the block encoding, $\kappa_{\brm{A}}$ is (an upper bound on) the effective condition number of $\brm{A}$, and $\varepsilon$ is the accuracy~\cite{Chakraborty2023quantumregularized}. Their algorithm combines QSVT and variable-time amplitude amplification (VTAA)~\cite{ambainis2012VTAA}, modifying the previous approach~\cite{chakraborty_et_al2018PowerBlockEncoding}. 

These algorithms do not lead to efficient quantum algorithms for $k$-HodgeRank.
Known block encodings of $\brm{B}_k$ are constructed via quantum random access memory (QRAM) or a sparse access input model (SAIM). In these cases, $\alpha_{\brm{B}_k}$ scales with either the Frobenius norm or the sparsity of $\brm{B}_k$~\cite{chakraborty_et_al2018PowerBlockEncoding,Chakraborty2023quantumregularized}, which both have worst-case scaling $\Omega(n^{k})$ as discussed previously. We summarize these relationships between classical least squares algorithms, quantum least squares algorithms, HodgeRank, and quantum $k$-HodgeRank in Figure~\ref{QLS comparison} in Supplemental Material~\cite{supp}.

We present a quantum algorithm to solve $k$-HodgeRank. Given a quantum state that encodes $\brm{s}^k$ and a (classically stored) list of edges in the clique complex, our algorithm outputs a quantum state that approximates a state proportional to Eq.~\eqref{higher HodgeRank solution} using QSVT~\cite{Gilyen2019QuantumArithmetics} and a projected unitary encoding (PUE) of $\brm{B}_k$. Its complexity has quadratically worse dependence on the effective condition number than Ref.~\cite{Chakraborty2023quantumregularized} but a simpler implementation. Moreover, the PUEs of $\bk$ and $\bkk$ have scaling factors $\sqrt{n}$~\cite{Kerenidis2022QuantumStates} or $n$~\cite{berry2023analyzing} depending on implementation, thus avoiding the potentially exponential complexity arising from QRAM/SAIM. While the algorithm presented in Ref.~\cite{Chakraborty2023quantumregularized} could possibly be adapted to work for PUEs while maintaining improved dependence on the effective condition number, we avoid this for simplicity.

\textit{Quantum HodgeRank.}---
We now give our quantum algorithm for $k$-HodgeRank (quantum $k$-HodgeRank) that scales exponentially better than its classical counterparts in $k$. It prepares a quantum state that encodes an approximate solution to Eq.~\eqref{higher HodgeRank solution} with a circuit depth that does not, in general, depend on the dimension $k$. Our algorithm is a classically controlled quantum circuit, as the structure is determined by $\mathcal{G}$.

We build quantum $k$-HodgeRank on the recently developed quantum algorithm for topological signal processing (QTSP)~\cite{Leditto2023topological} for analyzing higher-order networks~\cite{Millan2020ExplosiveComplexes,Battiston2020NetworksDynamics,Battiston2021TheSystems,Bianconi_2021,Majhi2022DynamicsReview,Bick2023WhatNetworks}. We assume access to a quantum state preparation unitary $\brm{U}_{\mathrm{prep}}$ that takes classical data $\brm{s}^{k}$ and prepares
\begin{eqnarray}\label{simplicialquantumtsignal}
        \lvert\brm{s}^k\rangle&:=&\frac{1}{\lVert \brm{s}^k\rVert_{2}}\sum_{i\in[n_k]} \,s^k_i\:\lvert\sigma_k^i\rangle,
\end{eqnarray}
where $\lvert\sigma_{k}^{i}\rangle$ is an $n$-qubit computational basis state representing $\sigma_{k}^{i}$ defined in~\cite{Lloyd2016QuantumData,berry2023analyzing} (given in Eq.~(\ref{eq:simplex-state-def}) in Supplemental Material~\cite{supp}). We discuss the complexities of our applications in terms of the number of calls to $\brm{U}_{\mathrm{prep}}$. 

QTSP combines the quantum singular value transformation (QSVT)~\cite{Gilyen2019QuantumArithmetics,Martyn2021GrandAlgorithms} and PUEs $\brm{U}_{\brm{\bk}}$ and $\brm{U}_{\brm{\bkk}}$ of the boundary matrices $\bk$ and $\bkk$~\cite{Kerenidis2022QuantumStates,McArdle2022AQubits} (see Supplemental Material~\cite{supp}). It prepares, upon successful postselection of the ancilla (which occurs with probability approximately $\mathcal{N}^2:=\big\|H\left(\bk/\sqrt{n},\bkkdag/\sqrt{n}\right) \lvert\brm{s}^k\rangle\big\|_{2}^2$), the \textit{filtered} state 
\begin{eqnarray}\label{filtered simplicial signal state}
    \lvert\brm{s}^{k}_{\mathrm{fil}}\rangle:=\frac{1}{\mathcal{N}}\,H\left(\frac{\bk}{\sqrt{n}},\frac{\bkkdag}{\sqrt{n}}\right) \lvert\brm{s}^k\rangle.
\end{eqnarray}
Here $H(x,y):[-1,1]^{2}\rightarrow[-1,1]$ is a sum of two real polynomials $p(x)$ and $q(y)$, each with degree at most $D$, applied to the singular values of the respective matrices, where $p$ and $q$ are either both even or one is zero. The polynomial $H(x,y)$ is chosen so that $\ket{\brm{s}^{k}_{\mathrm{fil}}}$ extracts specific information from $\ket{\brm{s}^k}$. QTSP requires one call to $\brm{U}_{\mathrm{prep}}$ as well as $O\big(D\big)$ calls to $\brm{U}_{\bk}$, $\brm{U}_{\bkk}$, their inverses, and $O(D)$ 
applications of $\mathrm{C}_{\Pi_{j}}\mathrm{NOT}$ for $j \in \{k-1, k, k+1\}$, where $\Pi_j$ is the projector onto the states corresponding to $j$-simplices~\cite{Metwalli2021FindingComputer,Akhalwaya2022TowardsComputers}, as well as post-selection on the correct ancilla state. 
The $\mathrm{C}_{\Pi_j}\mathrm{NOT}$ gates depend on the edges in the underlying graph $\mathcal{G}$ of the clique complex $\mathcal{K}_n(\mathcal{G})$ and are implemented via classical control. The circuit for QTSP has non-Clifford gate depth $O(Dn\log n)$ and uses $O(n)$ qubits. We review QTSP and its complexity in Supplemental Material~\cite{supp} (see~\cite{Leditto2023topological}). 

Quantum $k$-HodgeRank utilizes QTSP to take as input the state $\ket{\brm{s}^k}$ as in Eq.~(\ref{simplicialquantumtsignal}) and output a quantum state $\ket{\tilde{\brm{s}}^{k-1}_{\ast}}$ that approximates 
\begin{eqnarray}\label{eqn:actualscore}
    \lvert\brm{s}^{k-1}_{\ast}\rangle:= \frac{1}{\mathcal{N}_{\ast}}\,\mathlarger{\mathlarger{\sum}}_{i\in[n_{k-1}]}\,{s}^{k-1}_{\ast,i}\,\lvert\sigma_{k-1}^{i}\rangle
\end{eqnarray}
where $\mathcal{N}_{\ast}:=\big\|\big(\bk\bk^{\dagger}\big)^{+}\bk\lvert\brm{s}^{k}\rangle\big\|_{2}$.
We apply an approximation of $\big(\bk\bkdag\big)^+\bk$, given by a polynomial $H\big(\bk/\sqrt{n},\bkkdag/\sqrt{n}\big) = p\big(\bk/\sqrt{n}\big)$ solely in $\bk$, to the input state. The polynomial $H(x,y)$ is defined by $q(y) = 0$ and $p(x)=xg_{\varepsilon}(x^2)$, where $\|g_{\varepsilon}(x) - 1/(2\kappa_{k}^{2}x)\| \leq (\varepsilon/(2\kappa_{k}^{2}))$ for all $x \in [-1,-1/\kappa_{k}^{2}]\cup[1/\kappa_{k}^{2},1]$ (see~\cite{Gilyen2019QuantumArithmetics,Hayakawa2022QuantumAnalysis}). Here, $\kappa_k$ satisfies $\kappa_k^{-1} \in(0,\sqrt{n}/\xi_{\mathrm{min}}^{(k)})$, where $\xi_{\mathrm{min}}^{(k)}$ is the spectral gap (smallest nonzero singular value) of $\brm{B}_k$, and $\varepsilon = (0,1/2)$. Importantly, $g_{\varepsilon}(x)$ is a polynomial of degree $O(\kappa_{k}^{2}\log(\sqrt{n}\kappa_{k}/\varepsilon))$, and 
\begin{eqnarray}\label{eqn:quantum-hodgerank-acc}
\left\|\left(\bk\bk^{\dagger}\right)^{+}\brm{B}_{k}
-\frac{2\kappa_{k}^{2}}{\sqrt{n}}\,p
\left(\frac{\bk}{\sqrt{n}}\right)\right\|_2\leq\varepsilon.
\end{eqnarray}
 
Let $\widetilde{\mathcal{N}}_* := \big\|H(\bk/\sqrt{n})\ket{\brm{s}^k}\big\|_2$. The output of quantum $k$-HodgeRank, after successful post-selection on the ancilla (which occurs with probability $\Omega\big(n\widetilde{\mathcal{N}}_*/(4\kappa_k^4)\big)$), is 
\begin{eqnarray}\label{quantum HodgeRank state}
    \lvert\tilde{\brm{s}}^{k-1}_{\ast}\rangle:=\frac{\sqrt{n}}{2\kappa_{k}^{2}\widetilde{\mathcal{N}}_{\ast}}\,\mathlarger{\mathlarger{\sum}}_{i\in[n_{k-1}]}\,\tilde{s}^{k-1}_{i,\ast}\,\lvert\sigma_{k-1}^{i}\rangle.
\end{eqnarray}
This is a $2\varepsilon/\left(\mathcal{N}_{\ast} - \varepsilon\right)$-approximation to $\lvert\brm{s}^{k-1}_{\ast}\rangle$ (in $\ell_2$ norm) if $\mathcal{N}_* > \varepsilon$. In cases where $\brm{s}^k$ has high rankability (defined formally in Application 1), $\varepsilon$ can be chosen such that $\| \lvert\tilde{\brm{s}}^{k-1}_{\ast}\rangle - \lvert{\brm{s}}^{k-1}_{\ast}\rangle\| \leq 4\varepsilon/\mathcal{N}_{\ast}$ without significantly affecting the complexity (see Supplemental Material~\cite{supp}).
The complexity of QTSP implies
that preparing $\ket{\tilde{\brm{s}}^{k-1}_*}$ from $\ket{\brm{s}^k}$, excluding post-selection, uses $O(n)$ qubits, a single call to $\brm{U}_\mathrm{prep}$, and has non-Clifford gate depth
\begin{eqnarray}\label{quantum HodgeRank complexity}
    O\big(n\kappa_{k}^{2}\log(n)\log(\sqrt{n}\kappa_{k}^{2}/\varepsilon)\big).
\end{eqnarray}
If $\kappa_k$ is small (e.g. not growing exponentially in $k$), this is much more efficient than classically computing $\brm{s}^{k-1}_*$, even for small $k$. Examples of simplicial complexes with slowly-vanishing (or even constant in $k$) spectral gap are given in~\cite{berry2023analyzing} and discussed in Supplemental Material~\cite{supp}. Computing $p(\bk)\brm{s}^k$ classically requires $\tilde{O}(\kappa_{k}^{2})$ iterations of matrix-vector multiplication that each require $O(n_k)$ operations~\cite{Yang2021FiniteComplexes}. This again highlights the dependency of the simplex dimension in the classical complexity, particularly for complexes with many $k$-simplices, i.e. when $n_k = \Theta\left(\binom{n}{k+1}\right) \approx \Theta(n^{k+1})$. 

Similar to other quantum  linear algebra algorithms~\cite{Harrow2009QuantumEquations,chakraborty_et_al2018PowerBlockEncoding}, the speedup of our algorithm comes from preparing the approximate solution state $\ket{\tilde{\brm{s}}^{k-1}_*}$. Extracting a precise classical representation of $\lvert\tilde{\brm{s}}^{k-1}_{\ast}\rangle$ is generally expensive. 
We present applications that recover relevant information while avoiding full-state tomography where possible. These applications use quantum $k$-HodgeRank (or a straightforward modification) as a subroutine, and their complexities do not, in general, depend on $k$. This indicates that quantum $k$-HodgeRank, and QTSP in general, can potentially speed up specific computations defined on high-dimensional simplicial complexes. We assume that $\kappa_{k}$ and $\kappa_{k+1}$ of $\bk$ and $\bkk$ are known and (ideally) small.

\textit{Approximating consistency measures.}---We approximately measure the rankability of data in terms of so-called consistency and local inconsistency measures $\mathrm{R}(k)\in [0,1]$ and $\mathrm{R}_\mathrm{C}(k)\in [0,1]$, generalising definitions from~\cite{Jiang2011StatisticalTheory}. 
The consistency measure associated to the data $\brm{s}^k$ is the (Euclidean) length of $\brm{s}^k_\mathrm{G}$ relative to $\brm{s}^k$. This is analogous to a least squares ``goodness of fit''. Similarly, the local inconsistency measure is the relative length of the projection of $\brm{s}^k$ onto $\mathrm{im}(\brm{B}_{k+1})$, which we denote $\brm{s}^{k}_\mathrm{C}$. Explicitly,
\begin{eqnarray}\label{reliability measures}
    \mathrm{R}(k) := \left( \frac{\brm{s}^k_\mathrm{G} \cdot \brm{s}^{k}}{\|\brm{s}^{k}\|^2}\right)^{1/2} \mbox{ and  } \mathrm{R}_\mathrm{C}(k) := \left(\frac{\brm{s}^k_\mathrm{C} \cdot \brm{s}^{k}}{\|\brm{s}^{k}\|^2}\right)^{1/2}.
\end{eqnarray}
The value $\mathrm{R}(k)$ quantifies whether (sets of) alternatives can be meaningfully ranked based on the data, where larger values indicate high rankability. Conversely, $\mathrm{R}_{\mathrm{C}}(k)$ measures the presence of non-transitive preference relationships in the data, where the consensus prefers $A$ over $B$, $B$ over $C$, but $C$ over $A$. This has applications in finding arbitrage in currency exchange markets~\cite{Jiang2011StatisticalTheory}. These measures are related to what is known as the Hodge decomposition (see Supplemental Material~\cite{supp} for a brief introduction to the underlying topology).

We provide algorithms to compute $\varepsilon$-approximations to $\mathrm{R}(k)$ and $\mathrm{R}_\mathrm{C}(k)$ (see Supplemental Material~\cite{supp} for details).
Eq.~\eqref{reliability measures} implies that $\mathrm{R}(k) = \left(\bra{\brm{s}^k} \Pi_{\mathrm{G}} \ket{\brm{s}^k}\right)^{1/2}$, where $\Pi_{\mathrm{G}} := \brm{B}_k^\dagger (\brm{B}_k \brm{B}_k^\dagger)^+ \brm{B}_k$ projects onto $\mathrm{im}(\bk^\dagger)$. Thus, we approximate $\mathrm{R}(k)$ using an accelerated Hadamard test~\cite{lin2022lecture} by using QTSP to build a PUE of (an approximate) $\Pi_{\mathrm{G}}$. Let $H(x,y) = p(x) = x^2 g_{\varepsilon^2/2}(x^2)$, where as before $g_\varepsilon$ is a polynomial which $\varepsilon/(2\kappa_k^2)$-approximates $1/(2\kappa_k^2 x)$. Then $\big\| \Pi_{\mathrm{G}} - 2\kappa_k^2 \,p\big(\bk/\sqrt{n}\big)\big\|_{2} \leq \varepsilon^{2}/2$, where $p(x)$ is degree $D=O\big(\kappa_{k}^{2}\log(n\kappa_{k}^{2}/\varepsilon^{2})\big)$ (see~\cite{Leditto2023topological}, Theorem 2). 
Thus, if $\brm{U}_{\Pi_\mathrm{G}} = \mathrm{QTSP}(k, \mathcal{K}_n, x^2 g_{\varepsilon^2/2}(x^2))$ is the unitary that block encodes $p\big(\bk/\sqrt{n}\big)$ in the subspace defined by the ancilla in state $\ket{0}^{\otimes a}$, then
\begin{eqnarray}
    2\kappa_{k}^{2}\bra{\brm{s}^k} \bra{0}^{\otimes a} \brm{U}_{\Pi_\mathrm{G}}\ket{\brm{s}^k} \ket{0}^{\otimes a} = \left(\mathrm{R}(k)\right)^2 \pm \varepsilon^2/2.
\end{eqnarray}
The value of $\mathrm{R}(k)$ can be $\varepsilon$-approximated with success probability $1-\delta$ using $O(\kappa_{k}^{2} \log(1/\delta)/\varepsilon^2)$ controlled applications of $\brm{U}_{\mathrm{prep}}$ and $\textproc{QTSP}\big(k,\mathcal{K}_{n},x^{2}g_{\varepsilon^2/2}(x)\big)$. 
Thus, if $\brm{U}_{\mathrm{prep}}$ has non-Clifford gate depth $G$, 
the overall non-Clifford gate depth of this procedure is
$O\left(\kappa_{k}^2 \varepsilon^{-2}\log(1/\delta) \left( G + n \kappa_k^2 \log n \log\left( \frac{n\kappa_k^2}{\varepsilon^2}\right) \right)\right) \approx \tilde{O}\left(\kappa_{k}^2 \varepsilon^{-2}\log(1/\delta) \left( G + n \kappa_k^2\right) \right)$, 
where $\tilde{O}$ hides logarithmic factors of $n$, $\varepsilon$, and $\kappa_k$. The total runtime of the analogous approximate computation via classical TSP~\cite{Barbarossa2020TopologicalComplexes,Schaub2021SignalBeyond} is $O\big( nkn_k\kappa_{k}^{2}\log(n\kappa_{k}^{2}/\varepsilon^{2})\big)$~\cite{Yang2021FiniteComplexes,Yang2022SimplicialFilters}, where $n_k \leq \binom{n}{k+1}$.

This application can serve as a quantum-accelerated litmus test to check whether performing $k$-HodgeRank classically is worthwhile without the expensive computation required to determine $\mathrm{R}(k)$ exactly. For example, knowing $\mathrm{R}(k)$ to additive error $\varepsilon \approx 0.1$ would be indicative of whether the underlying data $\brm{s}^k$ is amenable to a global ranking via $k$-HodgeRank. 

Computing the local inconsistency measure is similar. In this task we estimate $\mathrm{R}_{\mathrm{C}}(k) = \left(\bra{\brm{s}^k} \Pi_{\mathrm{C}} \ket{\brm{s}^k}\right)^{1/2}$, where $\Pi_{\mathrm{C}}$ projects $\mathrm{im}(\bkk)$. Instead of applying $p(x)$ as defined above, apply $H(x,y)=q(y)=y^{2}g_{\varepsilon^2/2}(y^2)$ such that $\|2\kappa_{k+1}^{2}\,q\big(\bkkdag/\sqrt{n}\big) - \Pi_{\mathrm{C}}\| \leq (\varepsilon^2/2)$. Analogously to approximating $\mathrm{R}(k)$, an accelerated Hadamard test can approximate $\mathrm{R}_\mathrm{C}(k)$ to additive error $\varepsilon$ with success probability $1-\delta$ using $\tilde{O}(\kappa_{k+1}^{2}\varepsilon^{-2} \log(1/\delta))$ applications of $\brm{U}_{\mathrm{prep}}$ and $\textproc{QTSP}\big(k,\mathcal{K}_{n},y^{2}g_{\varepsilon^2/2}(y^2)\big)$, with overall non-Clifford gate depth $\tilde{O}\left( \kappa_{k+1}^2 \varepsilon^{-2} \log(1/\delta) \left( G + n\kappa_{k+1}^2 \right) \right)$. The exact classical computation for this quantity, as given by Jiang et al.~\cite{Jiang2011StatisticalTheory}, is $O(n^9)$ in the $k=1$ case, up to cubically worse in $n$. Implementing this polynomially classically has complexity $O\big( nkn_{k+1}\kappa_{k+1}^{2}\log(n\kappa_{k+1}^{2}/\varepsilon^{2})\big)$ for general $k$, with the same exponential scaling pitfalls as computing $\mathrm{R}(k)$ classically. 

\textit{Relative rankings of few alternatives.}---We can use quantum $k$-HodgeRank to approximate the scores of individual alternatives. This gives a heuristic for relatively ranking a small subset of alternatives without full-state tomography or computing classical $k$-HodgeRank. For more details see Supplemental Material~\cite{supp}. 

Let $\brm{U}_{k-1}^{i}$ be the tensor product of $\brm{X}$ gates such that $\brm{U}_{k-1}^{i}\ket{0\dots0} = \ket{\sigma_{k-1}^i}$. By calling controlled applications of quantum $k$-HodgeRank, $\brm{U}_{\mathrm{prep}}$, and $\brm{U}_{k-1}^{i}$, a Hadamard test can approximate the score $s^{k-1}_{*, i}$ of each $(k-1)$-simplex $\sigma_{k-1}^i$.
Approximating the scores for $L$ alternatives up to additive error $\varepsilon$ requires $O(n)$ qubits and $O\big(L\kappa_k^2\,\log(L/\delta)/(\sqrt{n}\varepsilon)\big)$ calls to each of $\textproc{QTSP}\big(k,\mathcal{K}_{n},xg_{\varepsilon/2}(x)\big)$, $\brm{U}_{\mathrm{prep}}$, and $\brm{U}_{k-1}^{i}$. Excluding the complexity of $\brm{U}_{\mathrm{prep}}$, this circuit has
non-Clifford gate depth is $\tilde{O}(L\kappa_k^{4}\sqrt{n}\log(L/\delta)/\varepsilon)$.
One can relatively rank $L$ alternatives $\sigma_{k-1}^{i_1},\dots, \sigma_{k-1}^{i_{L}} \in \mathcal{K}_n(\mathcal{G})$ by approximating each score individually and ranking them according to these approximations. 

This procedure is, in principle, efficient for small $L$. However, the number of $(k-1)$-simplices (i.e. basis states) is $O(n^{k})$, and so magnitude of the average amplitude may shrink exponentially in $k$. Thus, the accuracy and precision of $x g_{\varepsilon/2}(x^2)$ and the Hadamard test may need to scale exponentially with $k$ to extract meaningful information from the approximate amplitudes, and a good choice of $\varepsilon$ cannot be determined a priori. While there may be specific choices of clique complex $\mathcal{K}_n(\mathcal{G})$ and data $\brm{s}^k$ where the quantum advantage is greater, this is not generically the case.
 
\textit{Finding good alternatives.}---
We now discuss determining an approximate score for every $(k-1)$-simplex. As mentioned previously, full-state tomography is expensive. However, the tomography techniques given in~\cite{vanApeldoorn2022QuantumUnitaries} returns a classically stored $\ell_\infty$-norm approximation to $\ket{\tilde{\brm{s}}^{k-1}_*}$, denoted $\hat{\brm{s}}^{k-1}_{*}$. As detailed in Supplemental Material~\cite{supp}, one can determine the score of every $(k-1)$-simplex, up to additive error $O(\varepsilon)$, in $O(\kappa_k^4 n \varepsilon^{-2})$ applications of $\mathrm{QTSP}(k, \mathcal{K}_n, x g_\varepsilon(x))$ and $\brm{U}_{\mathrm{prep}}$.
This $\ell_\infty$-norm approximation can be used to find alternatives with scores within additive error $O(\varepsilon)$ of the best alternative, or to determine whether the alternatives can be classified into subsets with similar scores within subsets and large differences in scores between subsets. However, like the previous application this has issues as the amplitudes of the state $\ket{\brm{s}^{k-1}_*}$ may be arbitrarily close, and an appropriate choice of $\varepsilon$ cannot be easily determined from $\ket{\brm{s}^k}$.

\textit{Dequantization.} --- Chia et al.~\cite{Chia2020DequantizedQSVT} and Gharibian and Le Gall~\cite{Gharibian2022DequantizingConjecture} have given different frameworks for dequantizing algorithms based on QSVT. In Supplemental Material~\cite{supp} we show that these frameworks do not generically lead to efficient classical dequantizations of quantum $k$-HodgeRank or the computation of consistency or local inconsistency scores (Application 1). The reasons for this are: (i) difficulties obtaining polylogarithmic-time sampling and query access to $\brm{B}_k$, and (ii) there exist simplicial complexes such that $\brm{B}_k$ has Frobenius norm $\Theta(n^k)$ and row sparsity $\Theta(n)$. We present a family of such complexes, previously studied by Berry et al.~\cite{berry2023analyzing} in the context of quantum topological data analysis (QTDA). Each of these issues leads to inefficient dequantized algorithms in these frameworks. Methods for dequantizing QTDA rely on the relationship to estimating the dimension of the null space of an operator and do not obviously apply to our algorithms. An interesting future direction would be determining to what extent QTSP and quantum $k$-HodgeRank can be dequantized.

\textit{Conclusions.} ---
Quantum $k$-HodgeRank is a straightforward combination of tools from QSVT and QTDA that approximates certain quantities much faster than classical algorithms for higher-order networks. 
HodgeRank offers an alternative solution to finding the stationary distribution of the random walk on the web page graph~\cite{Jiang2011StatisticalTheory}, and so quantum HodgeRank could be considered a variant of quantum PageRank algorithms~\cite{Garneron2012Adiabatic,Sanchez-Burillo2012}. We note that while recommendation systems are related to rank aggregation problems, the problem formulation and methods used in HodgeRank differ significantly from those studied by Kerenidis and Prakash~\cite{Kerenidis2017QuantumSystem}. This distinction extends to the respective quantum algorithms. 

Comparing to a broader range of classical algorithms is required to determine the presence of quantum advantage. Approximate methods for computing $\brm{s}_*^{k-1}$, given in Eq.~(\ref{higher HodgeRank solution}), exist for $k=1$ and some limited cases for $k=2$ \cite{Cohen2014Solving,Black2022HodgeComplexes,Ding2022HardnessGadgets}. However, these methods do not extend to Laplacian linear equations for $k\geq2$, which are as hard to (approximately) solve as general linear equations~\cite{Ding2022HardnessGadgets}, where sparse approximate solvers for linear equations with $N$ nonzero coefficients and condition number $\kappa$ run in time $O(\min\left\{ N^{2.27159}, N\kappa\right\})$~\cite{Ding2022HardnessGadgets}. In the case of Laplacian {equations, $N = \Omega(n_k)$ can grow exponentially in $k$. This intuitively suggests that Equation~\eqref{higher HodgeRank solution} is hard to solve, even approximately.

The proof in~\cite{Ding2022HardnessGadgets} maps general linear equations to linear equations in the boundary matrix of a two-dimensional complex. This allows any least squares problem to be realised as a least squares problem in a boundary matrix $\brm{B}_2$. Thus, theoretically quantum $2$-HodgeRank can solve general linear equations. Practically, the complexity of the mapping is linear in the number of variables and so this is inefficient. However, generalizing their methods to arbitrary $k$-dimensional complexes may offer a new avenue for efficient quantum linear systems solvers. A simpler intermediate step would be to characterize the boundary matrices of simplicial or clique complexes, in order to determine the class of least squares problems for which quantum $k$-HodgeRank could offer a speed-up compared to classical algorithms or QRAM/SAIM-based quantum methods.

Broadening the range of applications of discrete exterior calculus to higher-order networks would give quantum $k$-HodgeRank (and QTSP) more concrete problems where quantum $k$-HodgeRank is likely to perform best. Potential applications to the Kuramoto model of coupled oscillators~\cite{Acebron2005KuramotoReview} and higher-order analogues~\cite{Millan2020ExplosiveComplexes,Carletti2023GlobalTopologicalSync}, as well as variants of boundary value problems in fluid dynamics~\cite{COTTER2014FEECFluids,Hirani2015FEECFluids,MOHAMED2016175FEECFluids} and plasma physics~\cite{Squire2012FEECVlasovMaxwell,Kraus_Kormann_Morrison_Sonnendrücker_2017GEMPIC} on discretized and triangulated domains. Applications with quantum data would circumvent the difficulties in preparing the initial state $\ket{\brm{s}^k}$. Possible avenues include measuring quantum contextuality~\cite{Abramsky_2011SheafContextuality,Abramsky2017ContextualFraction} or studying multipartite interactions in complex quantum networks~\cite{Nokkala2024ComplexQNetworks}.

\section*{Acknowledgements}
We thank Pedro C. S. Costa for several technical discussions. C.M.G.L. is supported by the Monash Graduate Scholarship and CSIRO Data 61 Top-Up Scholarship. 

\bibliographystyle{unsrt}
\bibliography{Main_QTSP}

\newpage
\onecolumngrid
\section{Review of Simplicial Complexes}\label{app:simplicial}

Here we give a brief overview of the necessary topology required to understand topological signal processing and $k$-HodgeRank; for a more detailed review of simplicial complexes and related algebraic topology see~\cite{Goldberg2002Complexes}.
A simplicial complex with node set $\mathcal{V} = \{v_1, \dots, v_n\}$, which we denote by $\mathcal{K}_n$ (sometimes dropping the $n$ subscript), is a set of subsets of the node/vertex set $\mathcal{V}$ (these subsets are called simplices) such that (i) $\{v_i\} \in \mathcal{K}_n$ for all $i \in \{1, \dots, n\}$, and (ii) if $\sigma \in \mathcal{K}_n$ and $\tau \subset \sigma$, then $\tau \in \mathcal{K}_n$ (i.e., every subset of a simplex is a simplex). A particular family of simplicial complexes that are relevant to this work is the family of clique complexes. A clique complex $\mathcal{K}_n(\mathcal{G})$, for some graph $\mathcal{G}$, is a simplicial complex with the same node set as $\mathcal{G}$ where the set of $k$-simplices is the set of $(k-1)$-cliques in $\mathcal{G}$. The structure of $\mathcal{K}_n(\mathcal{G})$ is thus entirely determined by the graph $\mathcal{G}$.

We give some basic terminology here that we use when discussing simplicial complexes. A simplex $\sigma \in \mathcal{K}_n$ is called a $k$-simplex if $|\sigma| = k+1$; the $+1$ follows from the idea that a $k$-dimensional simplex in the geometric sense has $k+1$ vertices, e.g. a 2-dimensional simplex is a triangle which has 3 vertices. A simplex $\tau$ is a $q$-face of a $k$-simplex $\sigma$ if $\tau \subset \sigma$ and $|\tau| = q+1$, for $q < k$. To construct the boundary operators, we also associate an orientation to each simplex, which corresponds to the parity of a permutation of its constituent elements. That is, each simplex has two orientations, which are called positive and negative. For a simplex $\sigma = \{v_{i_0}, \dots, v_{i_k}\}$, we denote the oriented simplex by $\left[ v_{i_0}, \dots, v_{i_k} \right]$. 

Simplicial complexes are a well-studied class of higher-order networks because of their extra topological structure. One particularly relevant construction from algebraic topology associated to simplicial complexes that mathematically underpins our work is the space of $k$-chains $\mathfrak{C}_k := \mathfrak{C}_k(\mathcal{K})$ of a simplicial complex $\mathcal{K}$, defined as formal sums of $k$-simplices in the complex.
It is sufficient for our purposes to consider the space of $k$-chains as (isomorphic to) a real vector space $\mathbb{R}^{n_k}$ where the basis is the set of $k$-simplices in $\mathcal{K}$, i.e., $\mathfrak{C}_k:=\big\{\sum_{i=0}^{n_k-1}c_i\sigma_k^i\mid c_i\in \mathbb{R}\big\}$, where $n_k$ is the number of $k$-simplices in $\mathcal{K}$. The isomorphism to $\mathbb{R}^{n_k}$ can be seen by mapping each simplex $\sigma_{k}^i$ to $\brm{e}_i$, where $\{\brm{e}_i\}_{i \in [n_k]}$ is the standard basis for $\mathbb{R}^{n_k}$. 

The different spaces $\mathfrak{C}_k(\mathcal{K})$ associated with different dimensions $k$ are related to each other by maps known as boundary operators.
The $k$-th boundary operator $\partial_k:\mathfrak{C}_k \to \mathfrak{C}_{k-1}$ is a linear map such that for $\sigma_k=[v_{i_0},\cdots,v_{i_k}]\in \mathfrak{C}_k$,
\begin{eqnarray}\label{eqn:boundop}
    \partial_k\sigma_k&:=&\sum_{j=0}^{k}\,(-1)^j\,[v_{i_0},\cdots,v_{i_{j-1}},v_{i_{j+1}}\cdots,v_{i_k}].
\end{eqnarray}
Here, the vertex $v_{i_j}$ is removed from each simplex in the sum. That is, $\partial_k$ maps a $k$-simplex to a formal sum of its $(k-1)$-faces, which are the boundary of the $k$-simplex in a topological sense. 

Since the boundary operator acts linearly on the space of $k$-chains, it can be represented by a matrix. 
Let $\bm{\mathrm{B}}_k:\mathbb{R}^{n_k} \to \mathbb{R}^{n_{k-1}}$ be the matrix representation of $\partial_k$, using the canonical bases of $k$-simplices and $(k-1)$-simplices for $\mathfrak{C}_{k}(\mathcal{K})$ and $\mathfrak{C}_{k-1}(\mathcal{K})$ respectively. 
The entries of $\bm{\mathrm{B}}_k$ are
\begin{align}\label{boundary matrix}
    \left(\bm{\mathrm{B}}_k\right)_{ij}=\begin{cases}
        1,\quad&\mbox{if $i\neq j$, $\sigma_{k-1}^i\subset\sigma_k^j$, $\epsilon(\sigma_{k-1}^i)\sim \epsilon(\sigma_k^j)$}\\
        -1,\quad&\mbox{if $i\neq j$, $\sigma_{k-1}^i\subset\sigma_k^j$, $\epsilon(\sigma_{k-1}^i)\nsim \epsilon(\sigma_k^j)$}\\
        0,\quad&\mbox{otherwise},
    \end{cases}
\end{align}
where $\epsilon(\sigma_{k-1}^i)\sim \epsilon(\sigma_k^j)$ means $\sigma_{k-1}^i$ and $\sigma_k^j$ have the same orientation and $\epsilon(\sigma_{k-1}^i)\nsim \epsilon(\sigma_k^j)$ means $\sigma_{k-1}^i$ and $\sigma_k^j$ have different orientations (again referring to~\cite{Goldberg2002Complexes} for the full definition). The $\pm 1$ entries account for the orientation of each $k$-simplex relative to its boundary $(k-1)$-faces. The adjoint (equivalently transpose since $\bk$ is real) of the boundary operator $\bk^\dagger$ is called the coboundary operator and naturally maps from $\mathbb{R}^{n_{k-1}}$ ($\mathfrak{C}_{k-1}(\mathcal{K})$) to $\mathbb{R}^{n_k}$ ($\mathfrak{C}_{k}(\mathcal{K})$). 

The boundary and coboundary operators are used to construct the Hodge Laplacian (matrix) $\bm{\mathrm{L}}_k$, defined as
\begin{eqnarray}\label{hodgelaplacian}
\bm{\mathrm{L}}_k:=\begin{cases}
\bm{\mathrm{B}}_{k+1}\bm{\mathrm{B}}_{k+1}^\dagger,\quad&\mbox{for $k=0$}\\
\bm{\mathrm{B}}_k^\dagger\bm{\mathrm{B}}_k+\bm{\mathrm{B}}_{k+1}\bm{\mathrm{B}}_{k+1}^\dagger,\quad&\mbox{for $1<k<n-2$}\\
\bm{\mathrm{B}}_k^\dagger\bm{\mathrm{B}}_k,\quad&\mbox{for $k=n-1$.}
\end{cases}
\end{eqnarray}
Furthermore, $\bm{\mathrm{L}}_k^\ell:=\bkdag\bk$ and $\bm{\mathrm{L}}_k^u:=\bkk\bkkdag$ are defined as lower and upper Hodge Laplacian operators, respectively. Hence, Eq.~\eqref{higher HodgeRank solution} can be expressed as $\brm{s}^{k-1}_* = \left(\brm{L}_{k-1}^u\right)^+\brm{B}_k \brm{s}^k$. More details of these operators can be found in Ref.~\cite{Goldberg2002Complexes}.

The boundary and Laplacian operators define what is known as the Hodge decompostion. This states that $\mathbb{R}^{n_k}$, the vector space of real numbers assigned to $k$-simplices (i.e. the space of $k$-chains), can be partitioned into three orthogonal components
\begin{align}\label{eqn:hodge decomposition}
    \mathbb{R}^{n_k} = \mathrm{im}\left( \bkdag\right) \oplus \mathrm{im}\left( \bkk \right) \oplus \mathrm{ker}\left( \hl \right).
\end{align}
These are respectively known as the gradient, curl, and harmonic components, generalizing the language used in low-dimensional vector calculus. For an arbitrary $\brm{s}^k \in \mathbb{R}^{n_k}$, we write $\brm{s}^k = \brm{s}^k_\mathrm{G} + \brm{s}^k_\mathrm{C} + \brm{s}^k_\mathrm{H}$ where $\brm{s}^k_\mathrm{G}$, $\brm{s}^k_\mathrm{C}$, and $\brm{s}^k_\mathrm{H}$ belong to the respective subspaces. The dimension of $\mathrm{ker}(\hl)$ is known as the $k$-th Betti number of the simplicial complex $\mathcal{K}_n$, and is well studied in classical and quantum topological data analysis.

\section{Topological signal processing (TSP) and details of quantum topological signal processing (QTSP)}\label{app:qtsp}

We utilize a newly developed framework from signal processing called topological signal processing (TSP)~\cite{Barbarossa2020TopologicalComplexes} to design our algorithms. TSP extends the concept of signals and signal processing from discretized time domains to more topologically nuanced spaces. In the TSP framework, the signal is referred to as a simplicial signal $\brm{s}^{k} \in \mathbb{R}^{n_k}$ and is defined as a $k$-chain (formally, a $k$-cochain, but for our purposes these are isomorphic) on a simplicial complex $\mathcal{K}_n$, effectively assigning a real number to each $k$-simplex. To perform filtering operations on these signals that respect the topology of the underlying space, TSP employs a filter operator described by polynomials of lower and upper Hodge Laplacian operators $\hll=\bk^{\dagger}\bk$ and $\hlu=\bkk\bkk^{\dagger}$~\cite{Yang2021FiniteComplexes,Yang2022SimplicialFilters}, respectively, called a simplicial filter:
\begin{eqnarray*}\label{simplicial filter}
    H\left(\hll,\hlu\right)&:=&\mathlarger{\mathlarger{\sum}}_{i_\ell=1}^{d_{\ell}}h_{i_\ell}^{\ell}\left(\hll\right)^{i_\ell}+\mathlarger{\mathlarger{\sum}}_{i_u=1}^{d_{u}}h_{i_u}^u\left(\hlu\right)^{i_u}-h_{0}\brm{I}.
\end{eqnarray*}

The aim of TSP is to design such a filter $H(x,y)$ so that the filtered simplicial signal $\brm{s}^{k}_{\mathrm{fil}} := H(\brm{L}_{k}^{\ell},\brm{L}_{k}^{u})\,\brm{s}^{k}$ extracts particular information from the original signal $\brm{s}^k$. The form of the filter is determined by its polynomial coefficients, called the simplicial filter coefficients, and depends on the task at hand. The problem of finding the coefficients is known as the filter design problem. Yang et al.~\cite{Yang2022SimplicialFilters} designed simplicial filters to project a simplicial signal $\brm{s}^{k}$ onto signals $\brm{B}_{k}^{\mathrm{T}}\brm{s}^{k-1}$ and $\brm{B}_{k+1}\brm{s}^{k+1}$, which are similar to gradient and curl projections that we discuss in Application 1. However, the polynomials they use to approximate the projections are quite different to what we employ in this paper, as they used least squares methods to find the filter coefficients. Nevertheless, this particular application helps illustrate the connections between the Hodge decomposition and $k$-HodgeRank.

Recently a quantum topological signal processing (QTSP) algorithm was developed for performing TSP filtering on quantum computers. Here we briefly review the general QTSP filtering algorithm, and refer the interested reader to~\cite{Leditto2023topological} for more details. For our purposes in this paper, we slightly generalize the definition of a simplicial filter given in their paper to accommodate the ``filter'' we implement in quantum $k$-HodgeRank. As in Lemma 1 of~\cite{Leditto2023topological}, define $H(x,y):[-1,1]^2 \to [-1,1]$ as a multivariable polynomial that is the sum of $p(x)$ and $q(y)$, each with degree at most $D$ such that both polynomials are either both even, both odd, or one is zero. The definite parity of $H$ ensures that $H\left(\brm{B}_k, \brm{B}_{k+1}^\dagger\right)$ is a map from $\mathbb{R}^{n_k}$ to $\mathbb{R}^{n_j}$ for some $j \in \{k-1, k, k+1\}$ (that is, the output simplicial signals all correspond to simplices of the same dimension). 

Instead of encoding each entry of the boundary matrices via QRAM/SAIM to implement block-encoding of $\bk$ and $\bkkdag$, a projected unitary encoding of $\bk$ and $\bkkdag$ is used. This is done using what is known as the fermionic Dirac boundary operator representation~\cite{Akhalwaya2022RepresentationOperator}, which we denote $\brm{D}$. Using its efficient application via a Clifford loader circuit as described in~\cite{Kerenidis2022QuantumStates}, the PUE of $\bk$ is given by
\begin{eqnarray}
    \Pi_{k-1}\brm{D}\Pi_{k}=\frac{\bk}{\sqrt{n}},
\end{eqnarray}
where $\Pi_{k}$ is a $k$-simplex identifying function, given in~\cite{Metwalli2021FindingComputer,berry2023analyzing,McArdle2022AQubits}, such that 
\begin{eqnarray*}
        \Pi_{k}\lvert\sigma_{k}^{i}\rangle\lvert0\rangle=\begin{cases}
            \lvert\sigma_{k}^{i}\rangle\lvert1\rangle,\quad&\mbox{if $\sigma_{k}^{i}\in\mathcal{K}_{n}$}\\
            \lvert\sigma_{k}^{i}\rangle\lvert0\rangle,\quad&\mbox{if $\sigma_{k}^{i}\notin\mathcal{K}_{n}$}.
        \end{cases}
\end{eqnarray*}
The quantum circuit for $D$ does not depend on the dimension of the simplices, or even the simplicial complex itself. On the other hand, $\Pi_k$ is a sequence of multi-controlled \textsc{CNOT} gates determined by the adjacency matrix of the underlying graph $\mathcal{G}$, along with inequality checks against predetermined values that depend on $k$. This is done via classical control without using QRAM/SAIM, as we assume the adjacency matrix is given classically. We refer to this computational model as a classically controlled quantum circuit, as defined in the main body. Then, the quantum singular value transformation (QSVT) and linear combination of unitaries (LCU) methods provide a way of constructing a PUE of the simplicial filter operator $H\left(\bk/\sqrt{n},\bkkdag/\sqrt{n}\right)$ given by
\begin{eqnarray}\label{our QTSP filter}
    p\left( \frac{\brm{B}_k}{\sqrt{n}}\right) + q\left( \frac{\bkkdag}{\sqrt{n}} \right)&:=&\mathlarger{\mathlarger{\sum}}_{i_\ell\in[d_{\ell}]}\mathfrak{h}_{i_\ell}^{\ell}\left(\dfrac{\bk}{\sqrt{n}}\right)^{i_\ell}+\mathlarger{\mathlarger{\sum}}_{i_u\in[d_{u}]}\mathfrak{h}_{i_u}^u\left(\dfrac{\bkkdag}{\sqrt{n}}\right)^{i_u}.
    \end{eqnarray}

This generalizes the definition in~\cite{Leditto2023topological}, which is standard in classical TSP literature, as we do not restrict the filter to be a summation of polynomials of $\hll$ and $\hlu$ (which is equivalent to assuming that both $i_\ell$ and $i_u$ are even). In~\cite{Leditto2023topological}, it is shown that one can construct the quantum circuit $\textproc{QTSP}\big(k,\mathcal{K}_{n},H(x,y)\big)$ using $a=n+6$ ancilla qubits such that
\begin{eqnarray}\label{QTSP Filtering Algorithm/Function}
       \textproc{QTSP}\big(k,\mathcal{K}_{n},H(x,y)\big)\big(\lvert\brm{s}^{k}\rangle\lvert0\rangle^{\otimes a}\big)= \lvert0\rangle^{\otimes a}\,H\left(\frac{\bk}{\sqrt{n}},\frac{\bkkdag}{\sqrt{n}}\right)\lvert\brm{s}^{k}\rangle+\lvert\perp\rangle,
   \end{eqnarray}
where $\lvert\perp\rangle$ is an un-normalized state satisfying $\big(\brm{1}^{n}\otimes\langle0\rvert^{\otimes a}\big)\lvert\perp\rangle=0$, and $\ket{\brm{s}^k}$ is as defined in Eq.~(\ref{simplicialquantumtsignal}). Viewed this way, $\textproc{QTSP}\big(k,\mathcal{K}_{n},H(x,y)\big)$ is an implementation of QSVT~\cite{Gilyen2019QuantumArithmetics} with a function $H(x,y)$ and PUEs of $\bk$ and $\bkkdag$~\cite{Kerenidis2022QuantumStates,McArdle2022AQubits}. Upon successful postselection, the output state given in Eq.~\eqref{filtered simplicial signal state} is obtained. This algorithm, excluding post selection, has non-Clifford gate depth $O(Dn\log(n))$ and uses $n$ system qubits (as described in Eq.~\eqref{eq:simplex-state-def}) and $O(n)$ ancilla qubits~\cite{Leditto2023topological}. 

To complete the definition of $\brm{s}^k$ given in Eq.~(\ref{simplicialquantumtsignal}), here we give the definition of $\ket{\sigma_k^i}$, as originally defined in~\cite{Lloyd2016QuantumData}. Recall that $\sigma_k^{i}$ is a $(k+1)$-element subset of $[n]$. Suppose that $\sigma_{k}^i = \left\{ v_{i_0}, \dots, v_{i_k} \right\}$. Then
\begin{align}\label{eq:simplex-state-def}
    \ket{\sigma_k^i} = \bigotimes_{j=1}^n \ket{u_j}, \quad u_j = 
    \begin{cases}
        1 & v_{u_j} \in \sigma_{k}^i\\
        0 & \text{otherwise.}
    \end{cases}
\end{align}
In other words, $\ket{\sigma_k^i}$ is an $n$-qubit computational basis state where the $j$-th qubit is 1 if and only if vertex $v_j$ is in the set $\sigma_{k}^i$. 

\section{More Details on Quantum $k$-HodgeRank}\label{app:quantumhodgerank}

\begin{figure}[h]
\includegraphics[scale=0.47]{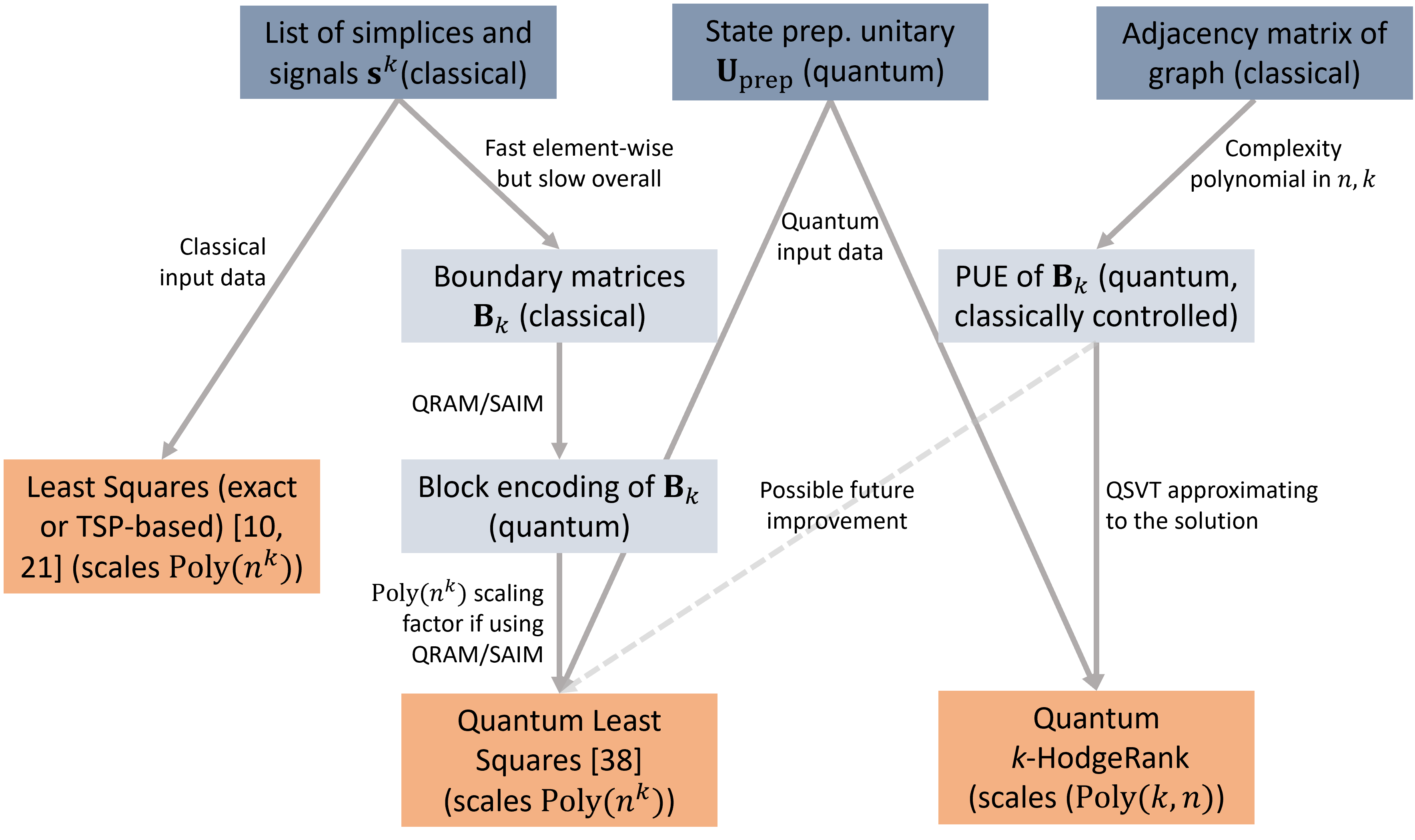}
\caption{\label{QLS comparison}
A comparison between several quantum least square (QLS) algorithms (with or without block-encoding (BE) method) and quantum $k$-HodgeRank to solve the $k$-HodgeRank problem given an input simplicial signal state $\ket{\brm{s}^{k}}$. Compared to classical methods, we do not need to explicitly construct the boundary operator element-wise, nor do we need to perform classical matrix-matrix or matrix-vector multiplication. Compared to general quantum least squares algorithms, such as those given by~\cite{Chakraborty2023quantumregularized}, quantum HodgeRank utilizes a method that constructs an operator that encodes the boundary matrix with a much smaller scaling factor, which leads to much faster computations if the clique complex has many $k$-simplices. While it is possible that the methods used in~\cite{Chakraborty2023quantumregularized} could be extended to utilize the PUE of $\brm{B}_k$, we do not explore that in this work.
}
\end{figure}

\subsection{Algorithm description and postselection probability}

Below we describe the problem our quantum $k$-HodgeRank algorithm aims to solve, excluding the state preparation unitary.

\begin{problem}\label{quantum HodgeRank problem}
    Given the simplicial signal state $\lvert\brm{s}^{k}\rangle$ that encodes the simplicial signal values $s_{i}^k$ for each $k$-simplex $\sigma_{k}^i$, output a quantum state that has amplitudes proportional to Eq.~(\ref{higher HodgeRank solution}).
\end{problem}
\noindent The quantum $k$-HodgeRank algorithm is essentially an implementation of QTSP. Quantum $k$-HodgeRank outputs, upon successful post-selection, a quantum state that encodes a normalized approximation to vector in Eq.~(\ref{higher HodgeRank solution}). 

First we review how to approximately implement the Moore-Penrose pseudoinverse of a matrix given by a projected unitary encoding. Let $\brm{U}_{\brm{A}}$ be a block-encoding or a projected unitary encoding of a matrix $\brm{A}$. Gilyen et al.~\cite{Gilyen2019QuantumArithmetics} apply QSVT to create a quantum circuit that encodes an approximation to $\brm{A}^+$, the Moore-Penrose pseudoinverse of $\brm{A}$. For simplicity, we use the slight modification of the statement as given in~\cite{Hayakawa2022QuantumAnalysis} (in the proof of their Theorem 5), which assumes $\brm{A}$ is Hermitian. Let $\xi_{\min}$ be the smallest non-zero singular value of $\brm{A}$ and let $\alpha$ be the scaling factor in the projected unitary encoding or block-encoding of $\brm{A}$. 
The idea is to implement QSVT with a polynomial $g_{\varepsilon}(x)$ of degree $O\big(\kappa\log(\kappa/(\alpha\varepsilon))\big)$, where $\kappa\in(\alpha/\xi_{\mathrm{min}},\infty)$ and $\varepsilon\in(0,1/2)$, such that (i) $\big|g_{\varepsilon}(x)\big|\leq1$ for all $x\in[-1,1]$, (ii) $\big|1/x-2\kappa g_{\varepsilon}(x)\big|\leq\varepsilon$ for $x\in[-1,-1/\kappa]\cup[1/\kappa,1]$, and (iii) $g_\varepsilon(0) = 0$. Applying $g_\varepsilon(x)$ to $\brm{A}$ approximately inverts every non-zero singular value while leaving the zero values untouched. Thus, it follows that $\big\|\brm{A}^+-2\kappa/\alpha\big(\brm{1}^{a}\otimes\bra{0}^{\otimes n}\big)\brm{U}_{\brm{A}}\big(\brm{1}^{a}\otimes\ket{0}^{\otimes n}\big)\big\|_2\leq\varepsilon$. 

In quantum $k$-HodgeRank, we want to implement $(\brm{B}_k \brm{B}_k^\dagger)^+\brm{B}_k$, given a PUE of $\bk$. This is naturally related to implementing the Moore-Penrose pseudoinverse of $\bk\bkdag$, which is a Hermitian matrix. It immediately follows that each eigenvalue of $\bk\bkdag$ is the square of the corresponding singular value of $\bk$. We use QSVT to implement the polynomial $xg_{\varepsilon}(x^2)$. Note that (i) since $|g_\varepsilon (x)| \leq 1$ for all $x \in [-1,1]$, it immediately follows that $|xg_{\varepsilon}(x^2)| \leq 1$ for all $x \in [-1,1]$, and (ii) if $g_{\varepsilon}(x)$ has degree $D$, then $xg_{\varepsilon}(x^2)$ has degree $2D+1$. This allows the tools developed in the proof of Theorem 5 in~\cite{Hayakawa2022QuantumAnalysis} to be applied even though $\brm{B}_k$ is not Hermitian (and indeed not even necessarily square). This is discussed briefly in~\cite{Leditto2023topological}. Recall the action of the QTSP circuit as described in Eq.~\eqref{QTSP Filtering Algorithm/Function}. The circuit to apply \textproc{QTSP}$\big(k,\mathcal{K}_{n},xg_{\varepsilon}(x^{2})\big)$ has non-Clifford gate depth $ O\big(n\kappa_{k}^{2}\log(n)\log(\sqrt{n}\kappa_{k}^{2}/\varepsilon)\big)$, as stated in Eq.~\eqref{quantum HodgeRank complexity}, which follows from the cost of QTSP given in Supplemental Material~\ref{app:qtsp} with $D = O\big(\kappa_{k}^{2}\log(\sqrt{n}\kappa_{k}^{2}/\varepsilon)\big)$. Recall that $\xi_{\mathrm{min}}^{(k)}$ is the smallest non-zero singular value of $\bk$ and let $\kappa_k \in (\alpha/\xi_{\min}^{(k)}, \infty)$ (in practice, one can choose $\kappa_k = \Theta\left(\sqrt{n}/\xi_{\min}^{(k)}\right)$. Problem~\ref{quantum HodgeRank problem} can be approximately solved as described in the following theorem.
\begin{theorem}\label{quantum HodgeRank}
    Fix $\varepsilon \in (0, \mathcal{N}_*)$. Upon correct postselection, the QTSP algorithm with input state $\ket{\brm{s}^k}$ and $H(x,y)=p(x)=xg_{\varepsilon}(x^{2})$ outputs a quantum state $\lvert\tilde{\brm{s}}^{k-1}_{\ast}\rangle$ that is $2\varepsilon/\left(\mathcal{N}_{\ast} - \varepsilon\right)$-close to $\lvert\brm{s}^{k-1}_{\ast}\rangle:=1/\mathcal{N}_{\ast}\,\big(\bk\bk^{\dagger}\big)^{+}\bk\lvert\brm{s}^{k}\rangle$ in $\ell_2$-norm. The cost of applying \textproc{QTSP}$\big(k,\mathcal{K}_{n},xg_{\varepsilon}(x^{2})\big)$ is $O\big(\kappa_{k}^{2}n\,\log(n)\log(\sqrt{n}\kappa_{k}^{2}/\varepsilon)\big)$ when $\mathcal{K}_n$ is a clique complex, and the probability of successful postselection is $O(n\mathcal{N}_{\ast}^{2}/\kappa_{k}^{4})$, where $\mathcal{N}_{\ast}:=\big\|\big(\bk\bk^{\dagger}\big)^{+}\bk\lvert\brm{s}^{k}\rangle\big\|_{2}$.
\end{theorem}
\begin{proof}
    It follows from~\cite{Leditto2023topological} (adapting Theorem 5 in~\cite{Hayakawa2022QuantumAnalysis}) that there exists a polynomial $p(x)=xg_{\varepsilon}(x^2)$ of degree $O\big(\kappa_{k}^{2}\log(\sqrt{n}\kappa_{k}^{2}/\varepsilon)\big)$ satisfying $\kappa_{k}\in(\sqrt{n}/\xi_{\mathrm{min}}^{(k)},\infty)$ and $\varepsilon\in(0,1/2)$, such that $\big|g_{\varepsilon}(x)\big|\leq1$ and $\left| x g_\varepsilon (x^2)\right|$ for all $x\in[-1,1]$ and $\big|1/x-2\kappa_{k}^{2}g_{\varepsilon}(x)\big|\leq\varepsilon$ for $x\in[-1,-1/\kappa_{k}^{2}]\cup[1/\kappa_{k}^{2},1]$. It follows that
    \begin{eqnarray}\label{eqn:poly-approx-1}
        \left\|\left(\brm{L}_{k-1}^{u}\right)^{+}\bk-\frac{2\kappa_{k}^{2}}{\sqrt{n}}p\big(\bk/\sqrt{n}\big)\right\|_{2}\leq\varepsilon,
    \end{eqnarray}
    recalling $\brm{L}^{u}_{k-1} = \brm{B}_k\brm{B}_k^\dagger$ as defined in Supplemental Material~\ref{app:simplicial}.
    Also note that $\ket{\brm{s}^{k-1}_*} = \frac{\left(\brm{L}_{k-1}^u\right)^+ \brm{B}_k \ket{\brm{s}^k}}{\|\left(\brm{L}_{k-1}^u\right)^+ \brm{B}_k \ket{\brm{s}^k}\|}$ is an alternative expression for $\ket{\brm{s}^{k-1}_*}$ as defined in Eq.~\eqref{eqn:actualscore}.
    Thus, we apply \textproc{QTSP}$\big(k,\mathcal{K}_{n},xg_{\varepsilon}(x^{2})\big)$ to the input state and the ancilla qubits such that
\begin{eqnarray}\label{eqn:polyapprox}
    \textproc{QTSP}\big(k,\mathcal{K}_{n},xg_{\varepsilon}(x^{2})\big)\big(\lvert\brm{s}^{k}\rangle\lvert0\rangle^{\otimes a}\big)=p\big(\bk/\sqrt{n}\big)\lvert\brm{s}^{k}\rangle \lvert0\rangle^{\otimes a}+\lvert\perp\rangle.
\end{eqnarray}
Upon the successful postselection on the ancilla qubits with probability $O(\mathcal{\widetilde{N}}_{\ast}^{2})$, where $\widetilde{\mathcal{N}}_{\ast}:=\big\|p\big(\bk/\sqrt{n}\big)\lvert\brm{s}^{k}\rangle\big\|_{2}$, the output is
\begin{eqnarray}\label{eqn:sk-1}
    \lvert\tilde{\brm{s}}^{k-1}_{\ast}\rangle=\frac{1}{\widetilde{\mathcal{N}}_{\ast}}\,p\left(\frac{\bk}{\sqrt{n}}\right)\lvert\brm{s}^{k}\rangle=\frac{1}{\widetilde{\mathcal{N}}_{\ast}}\,\mathlarger{\mathlarger{\sum}}_{i\in[n_{k-1}]}\,\frac{\tilde{s}^{k-1}_{\ast,i}}{2\kappa_{k}^{2}/\sqrt{n}}\,\lvert\sigma_{k-1}^{i}\rangle.
\end{eqnarray}
Eq.~\eqref{eqn:polyapprox} and the assumption that $\mathcal{N}_* > \varepsilon$ together imply that $\widetilde{\mathcal{N}}_* \geq \sqrt{n}\left(\mathcal{N}_{\ast}-\varepsilon\right)/(2\kappa_{k}^{2})$, as the reverse triangle inequality implies that
\begin{align*}
    \left\| p\left(\frac{\bk}{\sqrt{n}}\right)\lvert\brm{s}^{k}\rangle \right\| = \left\| p\left(\frac{\bk}{\sqrt{n}}\right)\lvert\brm{s}^{k}\rangle - \frac{\sqrt{n}}{2\kappa_k^2} \left(\brm{L}_{k-1}^{u}\right)^{+}\bk \ket{\brm{s}^k} + \frac{\sqrt{n}}{2\kappa_k^2} \left(\brm{L}_{k-1}^{u}\right)^{+}\bk \ket{\brm{s}^k} \right\| \geq \frac{\sqrt{n}}{2\kappa_k^2} \left| \mathcal{N}_* - x \right|
\end{align*}
for some $x \in (0,\varepsilon)$, which implies the desired lower bound on $\widetilde{\mathcal{N}}_*$. A similar argument implies the corresponding upper bound of $\widetilde{\mathcal{N}}_* \leq \sqrt{n}(\mathcal{N}_* + \varepsilon) / (2\kappa_k^2)$. It follows that
\begin{align*}
    \left\|\lvert\brm{s}^{k-1}_{\ast}\rangle-\lvert\tilde{\brm{s}}^{k-1}_{\ast}\rangle\right\|_{2}&= \left\|\frac{\big(\brm{L}_{k-1}^{u}\big)^{+}\bk\lvert\brm{s}^{k}\rangle}{\mathcal{N_{\ast}}}-\frac{p\big(\bk/\sqrt{n}\big)\lvert\brm{s}^{k}\rangle}{\mathcal{\widetilde{N}_{\ast}}}\right\|_{2}\\
    &= \frac{1}{\mathcal{N}_* \widetilde{\mathcal{N}}_*} \left\| \frac{\sqrt{n}}{2\kappa_k^2} \widetilde{\mathcal{N}}_* \left(\brm{L}_{k-1}^{u}\right)^{+} \brm{B}_k \ket{\brm{s}^k} - \mathcal{N}_* p\left( \brm{B}_k / \sqrt{n}\right) \ket{\brm{s}^k} \right\|\\
    &\leq \frac{1}{\widetilde{\mathcal{N}}_*} \left[ \left\| \frac{\sqrt{n}}{2\kappa_k^2} \left( \brm{L}_{k-1}^u \right)^+ \brm{B}_k \ket{\brm{s}^k} - p\left( \brm{B}_k /\sqrt{n}\right) \ket{\brm{s}^k}\right\| + \frac{\sqrt{n}}{2\kappa_k^2} \varepsilon\right]\\
    &\leq \frac{1}{\mathcal{N}_* - \varepsilon} \left[ \left\| \left( \brm{L}_{k-1}^u \right)^+\brm{B}_k \ket{\brm{s}^k}- \frac{2\kappa_k^2}{\sqrt{n}} p\left( \brm{B}_k / \sqrt{n} \right) \ket{\brm{s}^k} \right\| + \varepsilon \right]\\
    &\leq \frac{2\varepsilon}{\mathcal{N}_* - \varepsilon}.
\end{align*}
This proves the first claim of the theorem. Now we turn to the postselection probability. 
The probability of successful postselection (that is, obtaining $\lvert\tilde{\brm{s}}^{k-1}_{\ast}\rangle$), is $\widetilde{\mathcal{N}}_*^2$. By the aforementioned bounds on $\widetilde{\mathcal{N}}_*$, if $\widetilde{\mathcal{N}}_* > \varepsilon$, this is $\Omega(n(\mathcal{N}_{\ast} - \varepsilon)^2/(2\kappa_{k}^{2})^{2})$.
\end{proof}

We remark that $\mathcal{N}_{*}$ can be arbitrarily small in general. However, if the consistency measure $\mathrm{R}(k)$ is bounded away from zero, one can choose $\varepsilon$ that is small enough so that $\big\|\lvert\brm{s}^{k-1}_{\ast}\rangle-\lvert\tilde{\brm{s}}^{k-1}_{\ast}\rangle\big\|_{2}$ does not blow up. Let $\gamma_{\mathrm{G}}$ be a lower bound for $\mathrm{R}(k)$ set such that the rankability of $\lvert\brm{s}^{k-1}_{\ast}\rangle$ is considered good (this would be application dependent, but we use $\gamma_\mathrm{G} = 0.1$ as an example). Then, for data $\brm{s}^k$ such that $\mathrm{R}(k) \geq \gamma_\mathrm{G}$, it follows that
\begin{eqnarray*}
    \gamma_{\mathrm{G}}&\leq&\big\|\bkdag\big(\brm{L}_{k-1}^{u}\big)^{+}\bk\lvert\brm{s}^{k}\rangle\big\|_2\leq\big\|\bkdag\big\|_2\mathcal{N}_{*}\leq\sqrt{n}\mathcal{N}_{*},
\end{eqnarray*}
since the largest singular value of $\brm{B}_k$ is at most $\sqrt{n}$~\cite{Horak2013SpectraComplexes}. By setting $\varepsilon = \gamma_{\ast}/2 := \gamma_{\mathrm{G}}/(2\sqrt{n})$, Theorem~\ref{quantum HodgeRank} ensures that quantum $k$-HodgeRank outputs a solution that is still $(4\varepsilon/\mathcal{N}_*)$-close to $\lvert\brm{s}^{k-1}_{\ast}\rangle$ in vector $2$-norm upon correct postselection. In general, setting $\varepsilon = \frac{\varepsilon'}{4} \gamma_{\ast}$, for a sufficiently small choice of $\varepsilon'$,  ensures that the output after postselection is $\varepsilon'$-close to $\ket{\brm{s}^{k-1}_*}$ in vector $2$-norm since $\mathcal{N}_* - \varepsilon \geq \frac{\gamma_\mathrm{G}}{\sqrt{n}}\left( 1 - \varepsilon'/4\right)$. Since the complexity of $\mathrm{QTSP}(k, \mathcal{K}_n, x g_\varepsilon(x^2))$ depends only logarithmically on $\varepsilon$, changing $\varepsilon$ by a factor of $\Omega(n^{-1/2})$ (if $\gamma_G = \Omega(1)$, for example) causes only a constant change in the overall complexity.

Moreover, if we want to boost the probability of successful postselection to be $O(1)$, we can implement $O(\kappa_{k}^{2}/(\sqrt{n}\gamma_{\ast})$ rounds of (fixed-point oblivious) amplitude amplification~\cite{Berry2014ObliviousAA}. Hence, the total non-Clifford gate depth of outputting $\lvert\tilde{\brm{s}}^{k-1}_{\ast}\rangle$ that is $O(\varepsilon/\mathcal{N}_*)$-close to $\lvert\brm{s}^{k-1}_{\ast}\rangle$ with probability above $2/3$ is $O(\kappa_{k}^{2}D\sqrt{n}\log(n)/\gamma_{\ast})$. 

To complement this result, we give a brief proof here that if $\brm{s}^k$ is drawn Haar randomly from the unit sphere $\mathbb{S}^{n_{k}-1}$ (that is, if the data vector is a uniformly random unit-norm vector) then the average values of $\mathrm{R}(k)$ and $\mathrm{R}_\mathrm{C}(k)$ are proportional to the dimensions of $\mathrm{im}\left(\bkdag\right)$ and $\mathrm{im}\left(\bkk\right)$ respectively. This suggests that post-selection for computing $\ket{\tilde{\brm{s}}^k_*}$, $\mathrm{R}(k)$, and $\mathrm{R}_\mathrm{C}(k)$ is, in an average case, reasonably efficient provided that the dimension of the corresponding subspace is not $o(n_k)$.
\begin{lemma}
    Let $\brm{s}^k \in \mathbb{R}^{n_k}$ be a Haar random unit vector. Then $\Exp{\|\Pi_\mathrm{P} \ket{\brm{s}^k}\|^2} = \frac{\mathrm{dim}(\mathrm{im}(\Pi_\mathrm{P})))}{n_k}$ for $\mathrm{P} \in \{\mathrm{G}, \mathrm{C}, \mathrm{H}\}$.
\end{lemma}
\begin{proof}
    Define $P_\mathrm{P} = \|\brm{s}^k_\mathrm{P}\|^2$, that is, the length of the projected vector squared, where $\brm{s}^k$ is a uniformly random vector on $\mathbb{S}^{n_k - 1}$. Linearity of expectation and Eq.~\eqref{eqn:hodge decomposition} imply that
    \begin{align*}
        \mathbb{E}\left[ P_\mathrm{G} + P_\mathrm{C} + P_\mathrm{H} \right] = \mathbb{E}\left[ \|\brm{s}^k\|^2\right] = 1.
    \end{align*}
    Let $t_{i}^k$ be the coefficients of $\brm{s}^k$ when written in the orthonormal basis of eigenvectors of $\hl$. By the symmetry of the sphere, $\Exp{(t^k_i)^2}$ is independent of $i$. Therefore, 
    \begin{align*}
        \Exp{\|\brm{s}^k\|^2} = \Exp{\sum_{i=1}^{n_k} (t^k_i)^2} = n_k\Exp{(t^k_1)^2}.
    \end{align*}
   Thus, it must follow that $\Exp{(t^k_1)^2} = 1/n_k$. Since $\Exp{P_{\mathrm{P}}} = \sum_{i=1}^{\mathrm{dim}(\mathrm{im}(\Pi_\mathrm{P}))} \Exp{(t_i^k)^2}$, the result follows. 
\end{proof}

\subsection{Assumptions on the weights $w_{ij}$ in the $\ell_2$ minimization}

Here we discuss the assumption we make about $w_{ij}$ in Eq.~\eqref{statistical ranking problem}. Our quantum $k$-HodgeRank algorithm assumes that $w_{ij} = 1$ if $[v_i, v_j] \in \mathcal{E}$ and is $0$ otherwise. This is equivalent to the condition that there exists some constant $c \in \mathbb{N}$ such that every pair of alternatives is either ranked by zero people or $c$ people. This condition is called a \textit{balanced} case in statistical ranking~\cite{Jiang2011StatisticalTheory}. Our quantum $k$-HodgeRank algorithm requires a similar assumption, that $\brm{s}^{k-1}_*$ is the minimal choice $\brm{x} \in \mathrm{im}(\brm{B}_k^\dagger)$ with respect to the standard inner product. The primary reason for this restriction is because the boundary and coboundary operators in terms of fermionic Dirac operators, as originally given by~\cite{Lloyd2016QuantumData,Akhalwaya2022TowardsComputers}, are adjoint with respect to the standard unweighted inner product. Generalizing this without using QRAM/SAIM would require modification of the Dirac operator $\brm{D}$ and would likely be nontrivial, but is an interesting avenue for further exploration

\section{Application 1: calculation of HodgeRank (in)consistency measures}\label{app:consistency}

Here we give more details on the approximation of the (generalized) consistency measure $\mathrm{R}(k)$ and local inconsistency measure $\mathrm{R}_\mathrm{C}(k)$, as defined in Eq.~\eqref{reliability measures}. Since the methods for computing these are nearly identical, we combine their discussion. Let $\Pi_\mathrm{P}$ be a projector onto a particular subspace of $\mathbb{R}^{n_k}$, viewed as the space of $k$-chains. Here $\mathrm{P}$ is a placeholder for either $\mathrm{G}$ (projecting onto $\mathrm{im}( \brm{B}_k^\dagger )$) or $\mathrm{C}$ (projecting onto $\mathrm{im}\left( \brm{B}_{k+1}\right)$). The orthogonality of the gradient, curl, and harmonic spaces given by the Hodge decomposition (Eq.~\eqref{eqn:hodge decomposition}) implies that
\begin{align*}
    \bra{\brm{s}^k} \Pi_\mathrm{P} \ket{\brm{s}^k} = \frac{1}{\|\brm{s}^k\|^2} \langle \brm{s}^k, \brm{s}^k_\mathrm{P}\rangle =  \frac{\|\brm{s}^k_\mathrm{P}\|^2}{\|\brm{s}^k\|^2}.
\end{align*}
Thus, if we can obtain an $\varepsilon$-approximation to the inner product $\bra{\brm{s}^k} \Pi_\mathrm{P} \ket{\brm{s}^k}$, this is a $\sqrt{\varepsilon}$-approximation to the relative length of the projection of $\brm{s}^k$ onto the desired subspace. When $\mathrm{P} = \mathrm{G}$, this is the consistency measure $\mathrm{R}(k)$, and when $\mathrm{P} = \mathrm{C}$ this is the local inconsistency measure $\mathrm{R}_\mathrm{C}(k)$.

Let $\brm{U}_{\Pi_{\mathrm{P}}}:=\mathrm{QTSP}\left(k, \mathcal{K}_n, \frac{1}{2\kappa_\mathrm{P}^2}\tilde{\Pi}_\mathrm{P}\right)$ be the projected unitary encoding of the $\varepsilon'$-approximation to the projector $\frac{1}{2\kappa_\mathrm{P}^2}\Pi_\mathrm{P}$ (defined in Application 1 in the main text), where $\kappa_\mathrm{G} = \kappa_k$ and $\kappa_\mathrm{C} = \kappa_{k+1}$ are defined accordingly. Then 
\begin{align*}
    \mathrm{QTSP}\left(k, \mathcal{K}_n, \frac{1}{2\kappa_\mathrm{P}^2}\tilde{\Pi}_\mathrm{P}\right) \ket{0}^{\otimes a} = \frac{1}{2\kappa_\mathrm{P}^2}\tilde{\Pi}_\mathrm{P} \ket{\brm{s}^k}\ket{0}^{\otimes a}+ \sum_{x \neq 0} \ket{\mathrm{Garb}_x}\ket{x},
\end{align*}
where $\| \Pi_\mathrm{P} - \tilde{\Pi}_\mathrm{P}\| \leq \varepsilon'$.
This implies that 
\begin{align*}
    \bra{\brm{s}^k} \bra{0}^{\otimes a}\mathrm{QTSP}\left(k, \mathcal{K}_n, \frac{1}{2\kappa_\mathrm{P}^2}\tilde{\Pi}_\mathrm{P}\right)\ket{\brm{s}^k} \ket{0}^{\otimes a} = \frac{1}{2\kappa_\mathrm{P}^2} \bra{\brm{s}^k} \tilde{\Pi}_\mathrm{P} \ket{\brm{s}^k}.
\end{align*}
Recall that if $\|\Pi - \tilde{\Pi}\| \leq \varepsilon'$, then $\left| \langle \Pi \ket{x}, \ket{y}\rangle - \langle \tilde{\Pi}\ket{x}, \ket{y}\rangle \right| \leq \varepsilon'$ for all unit vectors $\ket{x}, \ket{y}$. Applying this with $\ket{x} = \ket{y} = \ket{\brm{s}^k}$ implies that $\bra{\brm{s}^k} \tilde{\Pi}_\mathrm{P} \ket{\brm{s}^k}$ is an $\varepsilon'$-approximation to $\bra{\brm{s}^k} \Pi_\mathrm{P} \ket{\brm{s}^k}$. 

Importantly, $\bra{\brm{s}^k} \tilde{\Pi}_\mathrm{P} \ket{\brm{s}^k}$ can be approximated via a Hadamard test to $\beta$ additive error with probability at least $1-\delta$ in $O(\log(1/\delta)/\beta^2)$ controlled applications of $\mathrm{QTSP}\left(k, \mathcal{K}_n, \frac{1}{2\kappa_\mathrm{P}^2}\tilde{\Pi}_\mathrm{P}\right)$ and $\brm{U}_{\mathrm{prep}}$. This can be improved to $O(\log(1/\delta)/\beta)$ applications using amplitude estimation, specifically by performing amplitude estimation on the Grover iterate that identifies the ancilla in state $\ket{0}^{\otimes a}$. The Grover iterate itself is straightforward to implement given access to controlled versions of the state preparation unitary $\brm{U}_\mathrm{prep}$. Knowing $\frac{1}{2\kappa_\mathrm{P}^2} \frac{\|\brm{s}^k_\mathrm{P}\|^2}{\|\brm{s}^k\|^2}$ to additive error $\beta$ corresponds to knowing $\frac{\|\brm{s}^k_\mathrm{P}\|}{\|\brm{s}^k\|}$ to additive error $\kappa_\mathrm{P}\sqrt{2\beta}$. 
Thus, using $O(\kappa_\mathrm{P}^2\log(1/\delta)/\varepsilon^2)$ applications of the QTSP implementation of the Hodge projector with approximation error $\varepsilon' = \frac{\varepsilon^2}{2}$, we can approximate $\frac{\|\brm{s}^k_\mathrm{P}\|^2}{\|\brm{s}^k\|^2}$ to error $\varepsilon$ with probability at least $1-\delta$. Combining this with the complexity for implementing $\textproc{QTSP}(k, \mathcal{K}_n, H(x,y))$ as given in Eq.~\eqref{quantum HodgeRank complexity} (or Theorem 2 in~\cite{Leditto2023topological}) gives the complexities stated in the main body in Application 1.

\section{Application 2: finding the relative ranking of alternatives}\label{app:relative}

\begin{problem}
   Let $\mathcal{M} = \big\{\sigma_{k-1}^{i}\big\}_{i\in[L]} \subset\mathcal{K}_{n}$ be a subset of $L$ distinct $(k-1)$-simplices in $\mathcal{K}_{n}$. Output a list of ordered scores $\big\{s^{k-1}_{*,i}\big\}_{i\in[L]}$ corresponding to such alternatives.   
\end{problem}
Given $\ket{\tilde{\brm{s}}^{k-1}_{\ast}}$, we can implement amplitude estimation to estimate the score assigned to each $(k-1)$-simplex by $\ket{\brm{s}^{k-1}_{\ast}}$. There are many different quantum amplitude estimation algorithms. Instead of aiming for optimality, we stick with a relatively simple Hadamard test to highlight the utility of repurposing tools from QTDA. 

Let $\brm{U}_{\mathrm{k-1}}^{i}$ be a state preparation unitary such that $\brm{U}_{\mathrm{k-1}}^{i}\ket{0}^{\otimes n}=\ket{\sigma_{k-1}^{i}}$ (which by Eq.~\eqref{eq:simplex-state-def} is simply parallel $\brm{X}$ gates on the appropriate system qubits) for $\sigma_{k-1}^{i}\in\mathcal{M}$ and define $\brm{U}_{\mathrm{amp}}^{i}:=\big(\brm{U}_{k-1}^{i}\otimes\brm{1}^{a}\big)^{\dagger}\textproc{QTSP}\big(k,\mathcal{K}_{n},xg_{\varepsilon'}(x^{2})\big)\big(\brm{U}_{\mathrm{prep}}\otimes\brm{1}^{a}\big)$ for some $\varepsilon' > 0$. Define a Hadamard test circuit
\begin{eqnarray}\label{Hadamard circuit}
    \mathrm{HAD}\big(\brm{U}\big):=\big(\brm{Had}\otimes\brm{1}^{ n}\otimes\brm{1}^{a}\big)\brm{c}\brm{U}\big(\brm{Had}\otimes\brm{1}^{n}\otimes\brm{1}^{a}\big),
\end{eqnarray}
where $\brm{cU}$ denotes a controlled application of $\brm{U}$. Then, we can estimate the rescaled amplitude $\big(\sqrt{n}/2\kappa_{k}^{2}\big)\,\tilde{s}^{k-1}_{\ast,i}=\bra{\sigma_{k-1}^{i}} p\left(\bk/\sqrt{n}\right)\ket{\brm{s}^{k}}=\bra{0}^{\otimes (n+a)}\brm{U}_{\mathrm{amp}}^{i}\ket{0}^{\otimes (n+a)}$
by estimating
$P(0)=\big(\bra{0}\otimes\brm{1}^{n+a}\big)\mathrm{HAD}\big(\brm{U}_{\mathrm{amp}}^{i}\big)\big(\ket{0}\ket{0}^{n+a}\big)$, the probability of a single shot of the Hadamard test returning the outcome $0$. The amplitude estimation requires $O(\log(1/\delta)/\varepsilon_{\mathrm{AE}})$ calls to $\mathrm{HAD}\big(\brm{U}_{\mathrm{amp}}^{i}\big)$ and its inverse, for some $\varepsilon_{\mathrm{AE}} > 0$, to obtain an estimate $\hat{s}^{k-1}_{\ast,i}$ that is $\big(2\kappa_{k}^{2}/\sqrt{n}\big)\varepsilon_{\mathrm{AE}}$-close to $\tilde{s}^{k-1}_{\ast,i}$ with probability of success $1-\delta$. This leads to
\begin{eqnarray*}
    \big|s^{k-1}_{\ast,i}-\hat{s}^{k-1}_{\ast,i}\big|&\leq&\big\|\big(\brm{L}_{k-1}^{u}\big)^{+}\bk\ket{\brm{s}^{k}}-\frac{2\kappa_{k}^{2}}{\sqrt{n}}p\left(\bk/\sqrt{n}\right)\ket{\brm{s}^{k}}\big\|_{2}+\big|\tilde{s}^{k-1}_{\ast,i}-\hat{s}^{k-1}_{\ast,i}\big|\\
    &\leq&\varepsilon'+\big(2\kappa_{k}^{2}/\sqrt{n}\big)\varepsilon_{\mathrm{AE}}=\varepsilon,
\end{eqnarray*}
by setting $\varepsilon'=\varepsilon/2$ and $\varepsilon_{\mathrm{AE}}=\varepsilon\sqrt{n}/(4\kappa_k^2)$. We perform this amplitude estimation procedure $L$ times, one for each $(k-1)$-simplex in $\mathcal{M}$, to obtain $\big\{\hat{s}^{k-1}_{\ast,i}\big\}_{i\in[L]}$. Given $\varepsilon,\delta>0$, the above application requires $O(n)$ qubits and $O\big(L\kappa_k^2\,\log(L/\delta)/(\sqrt{n}\varepsilon)\big)$ calls to each of $\textproc{QTSP}\big(k,\mathcal{K}_{n},xg_{\varepsilon/2}(x)\big)$, $\brm{U}_{\mathrm{prep}}$, and $\brm{U}_{k-1}^{i}$ to obtain an $\varepsilon$-approximation to each of the $L$ scores with probability $1-\delta$ (by the union bound over all $L$ procedures). The last step is then to order the scores to create a relative ranking of the elements in $\mathcal{M}$.

This is only a heuristic because, given some data $\brm{s}^k$, there is no clear way to choose $\varepsilon$ a priori to guarantee that the relative ranking of $\mathcal{M}$ obtained by the above procedure is the same as the relative ranking of these simplices according to $\brm{s}^{k-1}_*$. If a pair of amplitudes in $\ket{\brm{s}^{k-1}_*}$ corresponding to basis states in $\mathcal{M}$ differ by less than $2\varepsilon$, these two basis states may be out of order in the relative ranking. For example, suppose the amplitudes of $\ket{\brm{s}^k}$ are given by the sequence $\left( \frac{3}{\sqrt{n_{k-1}}}, \frac{2}{\sqrt{n_k-1}}, \frac{1}{\sqrt{n_{k-1}}}, \dots\right)$, and the aim is to relatively rank the first three alternatives. If $n_{k-1} \approx \binom{n}{k}$, then $\varepsilon = O\left( \binom{n}{k}^{-1/2} \right)$ is needed in order to deduce anything meaningful from the output. For this reason this application will not achieve a large (exponential), general speedup over the classical computation, even if it does well in some cases.

\section{Application 3: finding a good alternative}\label{app:good alternative}
\begin{problem}
    Given $\ket{\brm{s}^{k}}$, with high probability output an ordered list of alternatives $\{\sigma_{k-1}^{i}\}$ according to $\{\hat{s}^{k-1}_{\ast,i}\}$ that are $\varepsilon$-close to their scores $\{s^{k-1}_{\ast,i}\}$.
\end{problem}

As mentioned in the main text, recovering approximate amplitudes of a quantum state is difficult in general. For this problem we use $\varepsilon$-$\ell_\infty$-norm tomography as described in~\cite{vanApeldoorn2022QuantumUnitaries}. We present two of their theorems and discuss their applicability to quantum $k$-HodgeRank.
\begin{theorem}[Corollary 17, Proposition 19~\cite{vanApeldoorn2022QuantumUnitaries}]\label{thm:tomography}
    Let $\varepsilon, \delta \in (0,1)$ and let $\ket{\psi} = \sum_{j \in [d]} \alpha_j \ket{j}$ be a quantum state with $\alpha_j \in \mathbb{C}$. Then
    $O\left( \log(d/\delta)/\varepsilon^2 \right)$ applications (in parallel) of a controlled state-preparation unitary for $\ket{\psi}$ suffice to compute an $\varepsilon$-$\ell_\infty$-norm estimate $\tilde{\alpha}$ of $\alpha$ with success probability at least $1-\delta$. On the other hand, if we only wish to learn $\tilde{\alpha}$ up to a global phase, it is sufficient to have $O\left( \log(d)\log(d/\delta)/\varepsilon^2 \right)$ copies of $\ket{\psi}$, with the ability to perform unitary operations on each copy before measurement. 
\end{theorem}
We apply these results with the unitary $\brm{U}_\varepsilon = \mathrm{QTSP}(k, \mathcal{K}_n, x g_\varepsilon (x^2)) (\brm{U}_{\mathrm{prep}}\otimes\brm{1}^{a})$ which prepares the state described in Eq.~\eqref{eqn:polyapprox}, which plays the role of $\ket{\psi}$.
The dimension of the vector $\alpha$, i.e., the amplitudes of the state $\brm{U}_\varepsilon \ket{0}^{\otimes (n+a)}$, is $2^{n+a}$, where $a = O(n)$ is the number of ancilla qubits used. Thus, in the language of Theorem~\ref{thm:tomography}, $d = 2^{O(n)}$. Eqs.~\eqref{eqn:poly-approx-1} and \eqref{eqn:sk-1} together imply that 
\begin{align*}
\left| \bra{x} \left( \brm{B}_k \brm{B}_k^\dagger\right)^+ \brm{B}_k \ket{\brm{s}^k} - \frac{2\kappa_k^2}{\sqrt{n}}\bra{x} p\left( \brm{B}^k / \sqrt{n} \right) \ket{\brm{s}^k}\right| \leq \varepsilon.
\end{align*}
Setting $\ket{x} = \ket{\sigma_{k-1}^i}$ implies that $\left| \tilde{s}^{k-1}_{*, i} - s^{k-1}_{*, i} \right| \leq \varepsilon$ for all $i \in [n_{k-1}]$. Now suppose we have a classical representation of $\ket{\hat{\brm{s}}^{k-1}_*}$, an $\varepsilon'$-$\ell_\infty$ norm approximation to $\ket{\tilde{\brm{s}}^{k-1}_{*}}$, where $\varepsilon' = \frac{\varepsilon\sqrt{n}}{2\kappa_k^2}$. Then for each $i \in [n_{k-1}]$,
\begin{align*}
    \left| \hat{s}^{k-1}_{*,i} - {s}^{k-1}_{*,i}\right| \leq \left| \hat{s}^{k-1}_{*,i} - \tilde{s}^{k-1}_{*,i}\right| + \left| \tilde{s}^{k-1}_{*,i} - {s}^{k-1}_{*,i}\right| \leq 2\varepsilon.
\end{align*}
Applying Theorem~\ref{thm:tomography} with $d = 2^{O(n)}$ and $\delta = 2^{-n}$, it follows that $O(\kappa_k^4 \varepsilon^{-2})$ parallel, controlled applications of $\brm{U}_\varepsilon$ (i.e., of $\mathrm{QTSP}(k, \mathcal{K}_n, x g_\varepsilon (x^2))$ and $\brm{U}_{\mathrm{prep}}$) are required to learn $\ket{\hat{\brm{s}}^{k-1}_*}$ with probability $1-2^{-n}$, or $O(\kappa_k^4 n \varepsilon^{-2})$ applications $\brm{U}_\varepsilon$ are required to learn $\ket{\hat{\brm{s}}^{k-1}_*}$ up to a global phase with probability $1-2^{-n}$. 

Knowing the value of $\brm{s}^{k-1}_*$ up to a global phase amounts to knowing the global order of alternatives but not knowing whether it is ranked best-to-worst or worst-to-best. In many cases, this would be possible to determine at a glance, by simply comparing the two extremes manually. In this case, one could get away with only using copies of $\ket{\brm{s}^{k-1}_*}$, rather than conditional applications of $\mathrm{QTSP}(k, \mathcal{K}_n, x g_\varepsilon (x^2))$ and $\brm{U}_{\mathrm{prep}}$. However, due to the concrete implementation of $\mathrm{QTSP}(k, \mathcal{K}_n, x g_\varepsilon (x^2))$, this might not offer a huge advantage unless controlled access to $\brm{U}_{\mathrm{prep}}$ is unavailable.

This method for determining a good alternative is only a heuristic for similar reasons that were discussed in Supplemental Material~\ref{app:relative} for calculating relative rankings: there is no way to determine an appropriate choice of $\varepsilon$, and there are states where $\varepsilon$ will have to be exponentially small in $k$ to draw meaningful conclusions from this procedure. If there was some guarantee that the scores of the alternatives all differ by at least some $\varepsilon>0$ that does not vanish exponentially, then one can recover the exact ranking according to $k$-HodgeRank efficiently (provided the standard assumptions about $\kappa_k$ being small). However, this is not a reasonable assumption in general. A more relaxed assumption may be that if a small set of alternatives ($k$-simplices) have much higher scores than the rest, classifying whether alternatives lie in this set or not may be efficient.  

\section{Dequantization roadblocks}\label{sec:dequant}

Quantum $k$-HodgeRank and QTSP more generally are based on applying QSVT to the PUE of the boundary operator $\brm{B}_k$. Chia et al.~(\cite{Chia2020DequantizedQSVT}, Theorems 5.1 and 6.2) and Gharibian and Le Gall~(\cite{Gharibian2022DequantizingConjecture}, Theorem 4.1) have given different methods of dequantizing algorithms based on QSVT. Here we discuss their results as well as two notable hurdles to applying these results to our algorithms: (i) the lack of efficient sampling access to the boundary operator and (ii) the lack of sparsity guarantees. Both of these issues lead to inefficient dequantized algorithms for quantum $k$-HodgeRank and for computing the consistency and local inconsistency measures $\mathrm{R}(k)$ and $\mathrm{R}_\mathrm{C}(k)$. We discuss each framework and their issues as well as a family of simplicial complexes, originally studied by Berry et al. in the context of QTDA~\cite{berry2023analyzing}, for which these issues are particularly evident.

\subsection{Dequantization via $\ell_2$-norm sampling}

First we discuss the dequantization framework of Chia et al.~\cite{Chia2020DequantizedQSVT}. We first summarize their definitions of sampling and query access and then quote the relevant theorem, Theorem 6.2, from their paper. We also discuss Theorem 5.1 from their paper (about approximating even singular value transformation), particularly in the context of computing $\mathrm{R}(k)$ and $\mathrm{R}_\mathrm{C}(k)$. However, its statement is dependent on many more definitions given in their paper and we instead refer readers to their paper for its full statement and context.

In their framework, one has sampling and query access to a vector $\brm{v} \in \mathbb{C}^n$, denoted by $\mathrm{SQ}(\brm{v})$, with complexity $\mathrm{sq}(\brm{v})$ if one can 
\begin{enumerate}[label=(\roman*)]
    \item query for entries of $\brm{v}$, 
    \item obtain independent samples of the indices $i\in [n]$ with probability $|v_i|^2/\|\brm{v}\|^2$,  
    \item query for $\|\brm{v}\|$,
\end{enumerate}
each in time at most $\mathrm{sq}(\brm{v})$. 
Similarly, sampling and query access to a matrix $\brm{A} \in \mathbb{C}^{m\times n}$, denoted by $\mathrm{SQ}(\brm{A})$ and with complexity $\mathrm{sq}(\brm{A})$, is defined by the ability to: 
\begin{enumerate}[label=(\roman*)]
    \item have sampling and query access to each of the rows $\brm{A}(i, \cdot)$ of $\brm{A}$,
    \item have sampling and query access to $\brm{a}$, the vector of row norms of $\brm{A}$, 
\end{enumerate}
both in time at most $\mathrm{sq}(\brm{A})$. 
Finally, one has oversampling and query access to a vector $\brm{v} \in \mathbb{C}^{n}$, denoted by $\mathrm{SQ}_\phi(\brm{v})$, if we have 
\begin{enumerate}[label=(\roman*)]
    \item query access to the entries of $\brm{v}$,
    \item sampling and query access $\mathrm{SQ}(\tilde{\brm{v}})$ to a vector $\tilde{\brm{v}} \in \mathbb{C}^n$ satisfying $\|\tilde{\brm{v}}\|^2 = \phi \|\brm{v}\|^2$ and $|\tilde{v}_i|^2 \geq |v_i|^2$ for all $i \in [n]$
\end{enumerate}
in time at most $\mathrm{sq}_\phi(\brm{v})$. With these definitions in mind, the following theorem about dequantizing the quantum singular value transformation is given. Theorem 5.1 from the same paper does not require that $\|A\|_\mathrm{F} = 1$, but instead the complexity of obtaining the required matrix sketches scales at least quadratically in $\|A\|_\mathrm{F}$ and quadratically in $\varepsilon^{-1}$. 

\begin{theorem}[\cite{Chia2020DequantizedQSVT}, Theorem 6.2]\label{thm:dequant-2}
    Suppose we are given a matrix $\brm{A} \in \mathbb{C}^{m\times n}$ satisfying $\|\brm{A}\|_\mathrm{F} = 1$ via oracles for $\mathrm{SQ}(\brm{A})$ and $\mathrm{SQ}(\brm{A}^\dagger)$ with $\mathrm{sq}(\brm{A}), \mathrm{sq}(\brm{A}^\dagger) = O(\log(mn))$, a vector $\brm{b} \in \mathbb{C}^n$ via $\mathrm{SQ}(\brm{b}) \in \mathbb{C}^n$ with $\|\brm{b}\|=1$ and $\mathrm{sq}(\brm{b}) = O(\log n)$, and a degree-$d$ polynomial $p(x)$ of parity $d$ such that $|p(x)| \leq 1$ for all $x \in [-1,1]$. Then with probability $1-\delta$, for $\varepsilon$ a sufficiently small constant, we can get $\mathrm{SQ}_\phi(\brm{v}) \in \mathbb{C}^n$ such that $\|\brm{v} - p(\brm{A})\brm{b}\| \leq \varepsilon \|p(\brm{A})\brm{b}\|$ in $\mathrm{poly} \left( d, \frac{1}{\|p(\brm{A})b\|}, \frac{1}{\varepsilon}, \frac{1}{\delta}, \log mn \right)$ time.
\end{theorem}

In our case, $\brm{A} = \brm{B}_k$, and so the first hurdle is obtaining efficient query access to the boundary operator and its conjugate transpose. Query access to $\brm{B}_k$ and $\brm{B}_k^\dagger$ can be made efficient by indexing the rows and columns by $k$- and $(k-1)$-simplices respectively, where the simplices are stored as (not necessarily contiguous) substrings of $123\dots n$. Querying for an element $(\brm{B}_k)_{ij}$ is then equivalent to checking whether $\sigma_{k-1}^i$ is a substring of $\sigma_{k}^j$ and, if so, determining the position of the missing element (which determines the sign by Eq.~\eqref{eqn:boundop}). This can be done in a single pass of both strings and thus can be done in $O(k)$ time, which is $O(\mathrm{poly}(\log n))$ for most applciations. 

However, sampling access to $\brm{B}_k$ is not straightforward, nor is sampling/querying the vector of row norms $\brm{a}$, which is defined by $a_j^2 = |\{ \sigma_{k} \in \mathcal{K} ~|~ \sigma^j_{k-1} \subset \sigma_k\}|$, the number of $k$-simplices that contain each $(k-1)$-simplex $\sigma^j_{k-1}$ in the complex $\mathcal{K}$. In principle, a $(k-1)$-simplex could be a subset of every $k$-simplex or none. The brute-force approach of checking all possible $k$-simplices is $O(n)$. Classical TDA often uses a data structure called a simplex tree to store face and coface relationships of a complex $\mathcal{K}$ efficiently~\cite{Boissonnat2012SimplexTree}. The simplex tree can be viewed as a trie (a prefix tree) on strings of the alphabet $[n]$, in lexicographical order, that correspond to simplices in a given complex. There is a bijection between the vertices of the simplex tree and the simplices in the complex, with each vertex labelled by its (lexicographically) last vertex (called $\mathrm{last}(\sigma)$ for a simplex $\sigma$), and the dimension of the simplex is equal to its depth in the tree minus one. While this data structure can speed up many operations on simplicial complexes, the complexity of sampling and enumerating cofaces of a $k$-simplex $\sigma$ with this data structure is $O(k\mathcal{T}^{>k}_{\mathrm{last}(\sigma)})$, where $\mathcal{T}^{>k}_{\mathrm{last}(\sigma)}$ is the number of nodes in the simplex tree with label $\mathrm{last}(\sigma)$ and depth at least $k+1$. This is not in general $O(\mathrm{poly}(\log mn))$, and in Section~\ref{sub:example} we give an example of a complex for which this scales exponentially in $k$. 

It is worth noting that sampling and query access to $\bkdag$ in $O(k)$ operations is straightforward. Query access is as defined above, and each of the rows of $\bkdag$ contain exactly $k+1$ nonzero entries, all equal to $\pm 1$, and thus the rows all have identical norm. Sampling access to each row can be obtained by taking the corresponding $k$-simplex and deleting an element in the $(k+1)$-element string uniformly at random. Chia et al.~(\cite{Chia2020DequantizedQSVT}, Remark 6.4) show that if one has sampling and query access to $\brm{A}$, then one can probabilistically simulate similar access (technically, oversampling and query access) to a matrix $\brm{B}$ such that $\|\brm{A}^\dagger - \brm{B}\| \leq \varepsilon \|\brm{A}\|$ with probability $1-\delta$. However, the complexity of doing so is $\tilde{O}\left( \frac{\|A\|^{28}_\mathrm{F}}{\|A\|^{28} \varepsilon^{22}}\log^3 \delta^{-1} \right)$, where $\|\cdot\|_\mathrm{F}$ is the Frobenius norm. Aside from the scaling in $\varepsilon$, the harsh scaling in the Frobenius norm can be problematic for reasons described below.

The second hurdle that appears in applying this framework is the scaling with respect to the Frobenius norm of $\brm{B}_k$. In Theorem 5.1 of their paper, the size of the matrix sketches to ensure sufficient accuracy and precision of the resulting operator scales at least quadratically in the Frobenius norm of $\brm{A}$~\cite{Chia2020DequantizedQSVT}. On the other hand, Theorem~\ref{thm:dequant-2} presumes that the matrix $\brm{A}$ has Frobenius norm $1$, and thus to apply this result we must substitute $\brm{A} = \brm{B}_k / \|\brm{B}_k\|_\mathrm{F}$. This naturally shrinks the singular values by a factor of $\|\bk\|_\mathrm{F}$. This has knock-on effects for the polynomials we implement to create the state $\ket{\tilde{\brm{s}}^k_{*}}$ or approximate $\mathrm{R}(k)$ and $\mathrm{R}_\mathrm{C}(k)$: the domain on which each polynomial needs to be an $\varepsilon$-approximation must now contain $\xi_{\min}(\brm{B}_k)/\|\bk\|_\mathrm{F}$ (where $\xi_{\min}$ is the smallest non-zero singular value), instead of just containing $\xi_{\min}(\brm{B}_k)/\sqrt{n}$. This increases the degree of the resulting polynomials by a factor of $\|\bk\|_\mathrm{F}^2/n$. There exist simplicial complexes (including clique complexes, as we discuss later) where the Frobenius norm of $\brm{B}_k$ where $\|\bk\|_\mathrm{F} = \Omega(n^k)$. In such cases, applying either \Cref{thm:dequant-2} (Theorem 6.2 from their paper) or Theorem 5.1 from their paper leads to a dequantized algorithm with complexity growing exponentially in $k$. 

\subsection{Dequantization via sparsity}

Gharibian and Le Gall give a different method for dequantizing algorithms based on the QSVT~\cite{Gharibian2022DequantizingConjecture}. Their method assumes similar but different notions of query access to the matrices and vectors in question, and has a complexity that depends primarily on the sparsity of the matrix and the degree of the polynomial applied. We summarize their main result (Theorem 4.1) and then discuss its limitations in our setting, which are similar to those discussed above.

In this framework, query access to a vector $\brm{v} \in \mathbb{C}^n$ is defined by the existence of an $O(\mathrm{poly}(\log n))$-time classical algorithm that on input $i \in [n]$ outputs entry $v_i$. One has $\zeta$-sampling access to $\brm{v}$, for some parameter $\zeta \in [0,1)$, if the following three conditions are satisfied:
\begin{enumerate}[label=(\roman*)]
    \item there exists a $O(\mathrm{poly}(\log n))$-time classical algorithm $\mathcal{Q}_{\brm{v}}$ that on input $i \in [n]$ outputs $v_i$,
    \item there exists an $O(\mathrm{poly}(\log n))$-time classical algorithm $\mathcal{SQ}_{\brm{v}}$ that samples from a probability distribution $p:[n] \to [0,1]$ such that 
    \begin{align*}
        p(j) \in \left[ (1-\zeta)\frac{|v_j|^2}{\|\brm{v}\|^2},(1+\zeta)\frac{|v_j|^2}{\|\brm{v}\|^2} \right]
    \end{align*}
    for all $j \in [n]$,
    \item we are given a real number $m$ satisfying $|m -\|\brm{v}\|| \leq \xi \|\brm{v}\|$.
\end{enumerate}
In this framework query access to an $s$-sparse matrix $\brm{A} \in \mathbb{C}^{m\times n}$ (that is, $s$ is the largest number of non-zero entries in a row or column of $\brm{A}$) is defined to be the existence of two $O(\mathrm{poly}(\log mn))$-time classical algorithms $\mathcal{Q}^{\mathrm{row}}_{\brm{A}}$ and $\mathcal{Q}^{\mathrm{col}}_{\brm{A}}$ such that
\begin{enumerate}[label=(\roman*)]
    \item on input $(i,\ell) \in [m]\times [s]$ the algorithm $\mathcal{Q}^{\mathrm{row}}_{\brm{A}}$ outputs the $\ell$-th nonzero entry in row $i$ of $\brm{A}$ if this row has at least $\ell$ nonzero entries, and outputs an error message otherwise;
    \item on input $(j,\ell) \in [n]\times [s]$ the algorithm $\mathcal{Q}^{\mathrm{col}}_{\brm{A}}$ outputs the $\ell$-th nonzero entry of column $j$ of $\brm{A}$ if this column has at least $\ell$ nonzero entries, and outputs an error message otherwise. 
\end{enumerate}

Then the problem $\mathrm{SVT}(s, \varepsilon, \xi)$ is defined as follows. The problem $\mathrm{SVT}(s, \varepsilon, \zeta)$ is very similar to estimating $\mathrm{R}(k)$ and $\mathrm{R}_\mathrm{C}(k)$ described in Application 1, which can be seen by setting $\brm{u} = \brm{v} = \brm{s}^k/\|\brm{s}^k\|$.

\begin{tabular}{l l}
    Problem $\mathrm{SVT}(s, \varepsilon, \zeta)$ & \\
    \hline
    Input: & query access to an $s$-sparse matrix $\brm{A}\in \mathbb{C}^{m\times n}$ with $\|\brm{A}\|\leq 1$\\
    \quad & query access to a vector $\brm{u} \in \mathbb{C}^n$ such that $\| \brm{u} \| \leq 1$\\
    \quad & $\zeta$-sampling access to a vector $\brm{v} \in \mathbb{C}^n$ such that $\|\brm{v}\|\leq 1$\\
    \quad &  an even polynomial $P \in \mathbb{R}[x]$ of degree $2d$ with $|P(x)| \leq 1$ for all $x \in [-1,1]$\\
    Output: &  an estimate $\hat{z} \in \mathbb{C}$ such that $|\hat{z} - \brm{v}^\dagger P(\sqrt{\brm{A}^\dagger \brm{A}}) \brm{u} | \leq \varepsilon$.\\
    \hline
\end{tabular}

The following theorem is shown by Gharibian and Le Gall.
\begin{theorem}[\cite{Gharibian2022DequantizingConjecture}, Theorem 4.1] \label{thm:dequant-3}
    For any $s\geq 2$ and any $\varepsilon \in (0,1]$, the problem $\mathrm{SVT}(s, \varepsilon, \zeta)$ can be solved classically with probability at least $1 - 1/\mathrm{poly}(n)$ in time $\tilde{O}(s^{2d}/\varepsilon^2)$ time for any $\zeta \leq \varepsilon/8$. 
\end{theorem}

The first problem is finding efficient algorithms $\mathcal{Q}^{\mathrm{row}}_{\bk}$ and $\mathcal{Q}^{\mathrm{col}}_{\bk}$. One of these is easy: $\mathcal{Q}^{\mathrm{col}}_{\bk}$ can be implemented in time $O(k)$ by identifying each column with its respective $k$-simplex, stored as a $(k+1)$-element substring of $123\dots n$, as discussed in the previous dequantization framework. On the other hand, it is unclear how to implement $\mathcal{Q}^{\mathrm{row}}_{\bk}$ in $O(\mathrm{poly}(\log n))$-time. The reason for this is similar to the issues of efficiently implementing $\mathrm{SQ}(\bk)$ in the framework of Chia et al.: determining the $\ell$-th nonzero entry of a given row (or whether it exists at all) requires one to be able to quickly evaluate which $k$-simplices contain a given $(k-1)$-simplex. Since there are potentially $\Theta(n)$ such $k$-simplices, the worst-case complexity of doing this naively is $O(n)$. We are not aware of faster methods for this, even when restricting to clique complexes. 

This also leads to the second issue with applying Theorem~\ref{thm:dequant-3} to compute $\mathrm{R}(k)$ and $\mathrm{R}_\mathrm{C}(k)$: the complexity of the resulting classical algorithm will still be poor if $\brm{B}_k$ is not sufficiently sparse. As mentioned above, the sparsity $s$ depends on the number of $k$-simplices that contain a given $(k-1)$-simplex, which can be linear in $n$. Furthermore, the degree $d$ of the polynomial that we implement to calculate these values depends on $\kappa_k \in (\sqrt{n}/\xi_{\min}^{(k)}, \infty)$: specifically, recall that the polynomial we apply to compute $\mathrm{R}(k)$ has degree $O\left(\kappa_k^2 \log\left(\frac{\kappa_k^2 n}{\varepsilon}\right)\right)$, where $\varepsilon$ is the approximation error in the polynomial. There exist clique complexes such that $s = n/\mathrm{poly}(\log n)$ where $\kappa_k = \Theta(\sqrt{n})$, and we give such an example in \Cref{sub:example}. Thus, the complexity of $\varepsilon$-approximating $\mathrm{R}(k)$ or $\mathrm{R}_\mathrm{C}(k)$ using the framework described in Theorem~\ref{thm:dequant-3} would be $\tilde{O}\left( s^{2d}/\varepsilon^2\right) = \tilde{O}\left(n^{2d}/\varepsilon^2\right)$ where $d = O\left(\kappa_k^2 \log\left(\frac{\kappa_k^2 n}{\varepsilon}\right)\right)$. Even when $\kappa = \mathrm{poly}(n)$, this is extremely slow and insufficient to achieve comparable complexity to the quantum algorithms. 

\subsection{A simplicial complex which leads to slow dequantized algorithms}\label{sub:example}

We spotlight an explicit construction of Berry et al.~\cite{berry2023analyzing} an infinite family of clique complexes for which the above mentioned classical dequantization methods fare particularly poorly. Let $K(m,k)$ (called $K'(m,k)$ in~\cite{berry2023analyzing}) be the family of clique complexes formed by taking the complete graph $K_k$, replacing each vertex by an independent set of size $m$ (with all-to-all connectivity between independent sets), and putting exactly one edge within each set of size $m$. As an abuse of notation, we conflate the clique complex $K(m,k)$ with its underlying graph. Here $n = mk$, we suppose that $k = \lfloor\log n \rfloor$, and we focus on the task of applying $\mathrm{QTSP}(k,K(m,k), x^2g_\varepsilon(x^2))$ and approximating $\mathrm{R}(k-1)$ given oracular access to the state $\ket{\brm{s}^{k-1}}$.

The complex $K(m,k)$ can be constructed recursively by the relation $K(m,k) = K(m,k-1) * K(m,1)$, where the $*$ operation represents the join of two simplicial complexes, the complex formed by the union $\sigma \cup \sigma'$ of all pairs of simplices $\sigma \in K(m,k-1)$ and $\sigma' \in K(m,1)$. Because of this structure, the spectrum of its Laplacian can be determined explicitly; most notably, $\xi_{\min}(\brm{L}_{k-1}) = 2$ (see Lemma 11 of~\cite{berry2023analyzing}). One can prove this by induction, starting with the base case of $K(m,1)$, where the graph consists of $m$ vertices and a single edge and thus the zeroth order Hodge Laplacian $\brm{L}_0$ is equal to 
\begin{align*}
    \brm{L}_0 = \brm{B}_1\brm{B}_1^\dagger = 
    \left(\begin{matrix}
        1 & -1 & 0 & 0 & \dots \\
        -1 & 1 & 0 & 0 & \dots \\
        0 & 0 & 0 & 0 & \dots\\
        \vdots & & & & \\
    \end{matrix}\right).
\end{align*}
This is an $m\times m$ Hermitian matrix which has eigenvalue $2$ with multiplicity $1$ and $0$ with multiplicity $n-1$. 
The induction step follows by applying Theorem 4.10 of Duval and Reiner~\cite{duval2002shifted} which states that the spectrum of $\brm{L}_{k-1}(X*Y)$, denoted $\mathrm{spec}_{k-1}(X*Y)$, is given by
\begin{align*}
    \mathrm{spec}_{k-1}(X*Y) = \bigcup_{i + j = k-2} \mathrm{spec}_i(X) + \mathrm{spec}_j(Y),
\end{align*}
where addition of sets here means $A+B = \left\{a+b\mid a\in A,b\in B\right\}$. It follows from the Hodge decomposition (see~\eqref{eqn:hodge decomposition}) that the nonzero eigenvalues of $\brm{L}_{k-1}$ are exactly the union of nonzero eigenvalues of $\brm{L}_{k-1}^\ell$ and $\brm{L}_{k-1}^u$ (see~\cite{Barbarossa2020TopologicalComplexes}, Proposition 1). Since neither $\brm{L}_{k-1}^\ell$ nor $\brm{L}_{k-1}^u$ are the zero matrix for this complex, and both are Hermitian, each of them have a nonzero eigenvalue. Thus, the spectral gaps of $\brm{L}_{k-1}^\ell$ and $\brm{L}_{k-1}^u$ are each at least $2$. Finally, since $\brm{L}_{k-1}^\ell = \brm{B}_{k-1}^\dagger \brm{B}_{k-1}$ and $\brm{L}_{k-1}^u = \bk\bkdag$, it follows that the smallest nonzero singular values of $\brm{B}_{k-1}$ and $\bk$ are each at least $\sqrt{2}$ for this family of complexes, for all values of $m \geq 2$ and $k \geq 1$. 

The clique complex $K(m,k)$ contains at least $m^k$ $(k-1)$-simplices (i.e. $k$-cliques) and at least $km^{k-1}$ $k$-simplices. Thus, $\|\brm{B}_{k-1}\|_\mathrm{F} = \Omega(k^2 m^{k-1})$ (which follows from \labelcref{boundary matrix}). This is $n^{\Theta(k)}$ when $k = \mathrm{polylog}(n)$, which means that a dequantized algorithm designed using the framework given by Chia et al.~\cite{Chia2020DequantizedQSVT} would have complexity exponential in $k$, even with polylogarithmic time sampling and query access to $\brm{B}_{k-1}$. Analogous claims about $\brm{B}_k$ imply similarly slow dequantized algorithms for computing $\mathrm{R}_\mathrm{C}(k)$. 

On the other hand, the number of $(k-1)$-simplices that contain a given $(k-2)$-simplex (i.e., the number of $k$-cliques containing a given $(k-1)$-clique) is $\Omega(m)$ for some $(k-2)$-simplices. To see this, simply consider a $(k-1)$-clique that contains exactly one vertex from $k-1$ of the $k$ subsets. Such a $(k-1)$-clique can be extended to a $k$-clique by adding in any vertex from the final set of $m$ vertices. Thus, the sparsity $s$ of $\brm{B}_{k-1}$ is lower bounded by $m = n/\mathrm{polylog}(n)$. Furthermore, since $\kappa_{k-1} \geq \sqrt{n}/2$, it follows that the complexity a dequantization algorithm based on Gharibian and Le Gall's method would have complexity $\tilde{O}\left( n^{\Omega(n)}\right)$.

The structure of this clique complex also implies that the complexity of searching its simplex tree for $(k-1)$-cliques containing a given $(k-2)$-clique is exponential in $k$. To see this, note that $K(m,k)$ contains at least $m^{k-1}$ $(k-1)$-simplices containing the vertex $n$. For all such simplices, the label associated to their corresponding vertex in the simplex tree will therefore be $n$ and it will be at level $k$ in the simplex tree. Thus, the complexity of locating the cofaces of each $(k-2)$ simplex is $O(km^{k-1})$, which for $m = n/\lfloor \log n \rfloor$ is $O(n^k)$~\cite{Boissonnat2012SimplexTree}.   

We reiterate that the methods outlined here are not the only possible ways to build dequantized algorithms, nor the only ways to store and access $\brm{B}_k$, and we do not prove general lower bounds on the complexities of dequantized algorithms or subroutines. Furthermore, the clique complex exemplified here has nontrivial symmetry, and so there may be even more efficient sampling and query access algorithms for $\brm{B}_k$ in this specific case. However, the issues discussed in this section put forward a case that the projected unitary encoding of $\brm{B}_k$ is hard to imitate classically, particularly when the underlying clique complex contains many $k$-simplices for some $k = \omega(1)$. This classical-quantum discrepancy comes in part because $\|\brm{B}_k\| = O(\sqrt{n})$ despite the fact that it can have exponentially (in $k$) large Frobenius norm and sparsity that is linear in $n$. This allows the scaling factor in the PUE to be polynomial in $n$ while causing poor scaling for the dequantization frameworks discussed above.

\end{document}